\documentclass[12pt]{article}

\RequirePackage[OT1]{fontenc}
\usepackage{amsthm,amsmath,natbib}
\RequirePackage[colorlinks,citecolor=blue,urlcolor=blue]{hyperref}
\usepackage{multirow}
\def\bi{\begin{itemize}}
\def\ei{\end{itemize}}
\def\be{\begin{equation}}
\def\ee{\end{equation}}

\def\rm#1{\mathrm{#1}}

\usepackage[english]{babel}
\usepackage{amsfonts}
\usepackage{amssymb}
\usepackage{graphicx}
\usepackage{enumerate}
\usepackage{bm}
\usepackage{caption}
\usepackage{subfig}

\newcommand{\omegaV}{\mbox{\boldmath \(\omega\)}}
\newcommand{\greekvec}[1]{\mbox{\boldmath$#1$}}
\newcommand{\IR}{\mathbb{R}}

\usepackage{bbm}

\title{\textbf{Exceedance-based nonlinear regression of tail dependence}}

\author
{ \\ \ \\ 
	\textbf{Linda Mhalla$^1$\footnote{\textit{Corresponding author: linda.mhalla@unige.ch}}, Thomas Opitz$^2$, Val\'erie Chavez-Demoulin$^3$}\\
	\normalsize{$^{1}$Geneva School of Economics and Management, Universit\'{e} de Gen\`{e}ve, 1205 Gen\`{e}ve, Switzerland }\\
	\normalsize{$^{2}$Biostatistics and Spatial Processes, INRA, 84914, Avignon, France}\\
		\normalsize{$^{3}$Faculty of Business and Economics (HEC), Universit\' {e} de Lausanne, 1015 Lausanne, Switzerland}\\
}
\date{\today}
\begin{document}
	\maketitle
\noindent \textbf{Abstract} The probability and structure of co-occurrences of extreme values in multivariate data may critically depend on auxiliary information provided by covariates. In this contribution, we develop a flexible generalized additive modeling framework based on high threshold exceedances for estimating covariate-dependent joint tail characteristics for regimes of asymptotic dependence and asymptotic independence. The framework is based on suitably defined marginal pretransformations and projections of the random vector along the directions of the unit simplex, which lead to convenient univariate representations of multivariate exceedances based on the exponential distribution. Good performance of our estimators of a nonparametrically designed influence of covariates on extremal coefficients and tail dependence coefficients are shown through a simulation study. We illustrate the usefulness of our modeling framework on a large dataset of nitrogen dioxide measurements recorded in France between 1999 and 2012, where we use the generalized additive framework for modeling marginal distributions and tail dependence in monthly maxima. Our results imply asymptotic independence of data observed at different stations, and we find that the estimated coefficients of tail dependence decrease as a function of spatial distance and show distinct patterns for different years and for different types of stations (traffic vs. background). 
	
\section{Introduction} 	
Modeling co-occurrence patterns of extreme values arising in multi-component systems is crucial for an accurate prediction of aggregated risks. The limiting dependence structures for extreme values do not present a simple parametric form, such as the Gaussian dependence arising from the multivariate central limit theorem, which has spawned an extensive literature covering a wide range of parametric to fully nonparametric dependence models. When asymptotic independence arises, the limit model has a simple form but does not indicate the rate of convergence to this limit, suggesting alternative joint tail representations should be used to model the residual dependence at the observed levels of the process. In this paper, we use threshold exceedance data, and we develop nonparametric modeling approaches suitable for characterizing asymptotic dependence and asymptotic independence when the strength of the dependence and more generally its shape may be governed by additional information given by covariate data. Only a few approaches exist in the current literature for modeling and estimating covariate influence on the joint tail structure. In a parametric framework, \cite{Mhalla_DC_CD} proposed integrating covariate information through the parameters indexing angular density models, but we want to avoid such strong parametric assumptions. In the asymptotic dependence case, a nonparametric approach based on a baseline density for the dependence, modified through a density ratio to obtain a set of different dependence models according to covariate information, was developed by \citet{deCarvalho.Davison.2014}. More recently, \citet{Mhalla2017} proposed very flexible modeling of the extremal dependence based on a generalized additive model with shape constraints using blockwise maxima data. In this paper, we extend their approach from maxima to threshold exceedance data, and we lift the restriction on asymptotic dependence by proposing a set of tools that work for both asymptotic dependence classes.

%{\color{red} I put here some texts from the next sections... du coup l'intro ressemble un peu a un fourre-tout :-) . We will have to complete and smooth the text + the aim of the paper/innovation + litterature review. The notations are:
%\begin{itemize}
%\item $\mathbf{X}^{*}$ for simple max-stable process (that is with unit Fr\'echet margins). 
%\item $Z$ unit Frechet random variable
%\item $\mathbf{\tilde{Z}}$ nonnegative random vector with Pareto margins 
%\item $\tilde{X}=\log{\tilde{Z}}$ unit exponentially ditributed
%\end{itemize}
%\CT{wouldn't it be better to simply use $\mathbf{Z}$ for the simple max-stable processes? \\ and for instance $Z^E$ for $1/Z$, which has exponential distribution? \\
%	and then use $X$, $X^P$ and $X^E=1/X$ for data with either unit Fr\'echet, unit Pareto or unit exponential marginal distributions, on which we will apply censoring to perform inference on tail properties? We can also use something like $X^{obs}$ to denote the observed data at the original marginal scale. }
%}

%\subsection{Characterizing asymptotic dependence}
Classical extreme value theory and practice are founded on max-stable processes, which are the only non-trivial limits arising from pointwise normalized maxima taken over independent replicates of a continuous stochastic process $\mathbf{X}=\{X(s)\}_{s \in \mathcal{S}}$, with $\mathcal{S}$ a set of indexes. When observed over a finite-dimensional set, the $d$-dimensional joint distribution $F_{\mathbf{Z}}$ of a max-stable random vector $\mathbf{Z}=(Z_1,\ldots,Z_d)$ is of multivariate extreme value type with marginal generalized extreme value distributions and a max-stable dependence structure \citep[Section 9.2]{deHaan_Book}. A handy representation of $F_{\mathbf{Z}}$ is obtained when the marginal distributions are transformed to a common unit Fr\'{e}chet distribution $F_{Z_i}(z)=\exp(-1/z)\, \mathbbm{1}_{[0,\infty)}(z)$. The multivariate max-stable distribution is then written as $F_{\mathbf{Z}}=\exp\{-V(\mathbf{z})\}\, \mathbbm{1}_{[0,\infty)^d}(\mathbf{z})$ with the \emph{exponent function} $V$ measuring the strength and form of the dependence; the max-stable process $\mathbf{Z}$ is then said to be simple. We will assume this property throughout, without loss of generality. The convergence of the extremal dependence of any distribution $F$ in the domain of attraction of $F_{\mathbf{Z}}$ is chacterized by the property of multivariate regular variation \citep{Resnick1987}:
\begin{equation}\label{eq:mrv}
t\{1-F(t\mathbf{z})\}\rightarrow V(\mathbf{z})=-\log F_{\mathbf{Z}}(\mathbf{z}), \qquad \mathbf{z}>\mathbf{0}, \qquad t\rightarrow\infty.
\end{equation}
Max-stable models are useful when data are asymptotically dependent. When this assumption does not hold, limiting models are of little practical use as the property~\eqref{eq:mrv} fails to distinguish between asymptotic independence and exact independence \citep{Heffernan2007}, and models for the joint tails based on hidden regular variation \citep{Resnick2002} are indispensable. 

Joint tail characterizations at the interface of asymptotic dependence and independence are often presented in a bivariate setup. In more than two dimensions, some pairs of components may be asymptotically dependent while others are not; for a deeper theoretical treatment, see \citet{deHaan.Zhou.2011}. We say here that a stochastic process is asymptotically independent if all of its pairs of components have this property; the only possible max-stable limit in this case is full independence. In view of our aim of modeling the joint tail decay while abstracting away from the univariate marginal distributions, we now suppose that the $d$-dimensional random vector $\mathbf{X}^P=(X^P_1,\ldots,X^P_d)$ is nonnegative with Pareto marginal distributions by applying a marginal probability integral transform to $\bm X$ if necessary. The use of unit Fr\'echet margins (denoted by $\mathbf{X}^F=(X^F_1,\ldots,X^F_d)$), which are tail equivalent to $X^P_i$, would yield the same representations in the following. 

%If $F$ is asymptotically dependent and lies in the maximum domain of attraction of a bivariate extreme value distribution with exponent measure $V$, we get $c_1=c_2=1/2$ and
%\begin{equation}
%\ell_1(z_1,z_2) = \left\lbrace w(1-w)\right\rbrace ^{-1/2} \left[ 1-V\left\lbrace (1-w)^{-1},w^{-1}\right\rbrace \right], \notag
%\end{equation}
%where $w=z_1/(z_1+z_2)$. \\

%\subsection{Characterizing asymptotic independence}

\noindent With asymptotic independence of $X_1^{P}$ and $X_2^{P}$, their dependence strength vanishes as we move further into the joint tail. A general flexible representation of the joint tail, leading to a broad class of models suitable for asymptotic independence, was introduced by \cite{ledford_1996,ledford_1997} for bivariate random vectors and then generalized to the multivariate setup by \citet{Wadsworth.Tawn.2013}. We denote by $\omegaV=(\omega_1,\ldots,\omega_d)$ a \emph{direction} (also called \emph{weight} or \emph{angle}) that is on the unit simplex $S_{d}=\{ u \in \mathbb{R}_{+}^{d} \mid u_1+\cdots+u_d=1 \}$ in $\mathbb{R}_{+}^d$. Any positive random vector $\mathbf{x}=(x_1,\ldots,x_d)>\mathbf{0}$ can be represented as $\mathbf{x}=(x^{\omega_1},\ldots,x^{\omega_d})=\mathbf{x}^{\boldsymbol{\omega}}$.
The joint tail representation is
\begin{equation}
\rm{Pr} \left(X^P_1 > x^{\omega_1}, \ldots, X^P_d >x^{\omega_d}\right) = \ell(x;\omegaV) x^{-\lambda_{\omegaV}}, \quad x\rightarrow +\infty, \label{AI_WT}
\end{equation}
where $\ell$ is a slowly varying function at infinity for any value of $\omegaV \in S_d$ held fixed. The function $\lambda_{\omegaV}\leq 1$ is called the angular dependence function \citep{wadsworth_2012,Wadsworth.Tawn.2013}. It must satisfy certain shape constraints and describes the decay rate along rays in direction $\omegaV$. In the case of asymptotic dependence, we have $\lambda_{\omegaV}\equiv1$ and $\ell(x;\omegaV)\not\rightarrow 0$ as $x\rightarrow\infty$. If \eqref{eq:mrv} holds, then $\ell(x;\omegaV)$ tends to a limit expressed through values of the exponent function $V$, which for $d=2$ is $1/x^{\omega_1}+1/x^{\omega_2}-V(1/x^{\omega_1},1/x^{\omega_2})$. In the bivariate case, $\lambda_{\omegaV}$ generalizes the coefficient of tail dependence $\eta$ introduced by \cite{ledford_1996} and the dependence measure $\overline{\chi}$ \citep{Coles1999}, where $\eta=(1+\overline{\chi})/2=1/\{2\lambda(1/2,1/2)\}$ characterizes the joint tail decay rate along the diagonal of the first hyperoctant. The coefficient $\eta$ can be defined in $d$ dimensions as $1/\{d\lambda(1/d,\ldots,1/d)\}$. In a similar way, the extremal coefficient $\theta\in[1,d]$ \citep{Schlather.Tawn.2003} is related to $V$ via $\theta=V(1,\ldots,1)$.
% the multivariate Pickands dependence function via $\theta=dA(1/d,\ldots,1/d)$. 

In the remainder of the paper, Section~\ref{sec:proj} introduces projections based on weighted maxima and minima of multivariate random vectors leading to convenient univariate representations with appealing distributional properties for characterizing dependence in the tail. In Section~\ref{sec:inference}, we develop nonparametric inference for such dependence based on a generalized additive modeling framework allowing the inclusion of covariate influence. The simulation study in Section~\ref{sec:simu} illustrates the good performance of our methods when the model is exact for the data but also when it represents an asymptotic approximation to the true data distribution. In the application presented in Section~\ref{sec:appli}, we use our new techniques to reveal the influence of spatial distance and time on the co-occurrence patterns of extreme values in a large dataset of French air pollution data. Conclusions with an outlook on future work are given in Section~\ref{sec:conclusion}.

\section{Max- and min-projections}
\label{sec:proj}
We define the notions of \emph{max-projection} and \emph{min-projection} of a vector $\mathbf{x}=(x_1,\ldots,x_d)\geq \mathbf{0}$ with respect to a weight vector $\omegaV=(\omega_1,\ldots,\omega_d)\in S_{d}$. The max-projection is given as $\max_{\omegaV}(\bm x)=\max_{j=1}^{d} \omega_jx_j$, and the min-projection is defined as $\min_{\omegaV}(\bm x)=\min_{j=1}^{d} x_j/\omega_j$. The link between the two projections is established through the inversion $\max_{\omegaV}(\bm x)=1/\min_{\omegaV}(\bm x)$ using the convention that $1/0=\infty$ and $1/\infty=0$. 
\subsection{Asymptotic dependence}
\label{sec:ad}
We assume that the random vector $\mathbf{X}^F$ with distribution function $F$ is in the max-domain of attraction of a simple max-stable process $\mathbf{Z}$. The exponent function $V$ in \eqref{eq:mrv} characterizing a max-stable random vector $\mathbf{Z}$ is positive, continuous, convex, and homogeneous of order $-1$ such that $V(t\mathbf{z})=t^{-1}V(\mathbf{z})$ for $t>0$. Exploiting the homogeneity of $V$, an alternative characterization of the extremal dependence is possible through the Pickands dependence function $A$ \citep{Pickands_1981}, where
\begin{equation}\label{eq:pickands}
V(\mathbf{z}) = \left( \dfrac{1}{z_1}+ \cdots+ \dfrac{1}{z_d}\right) A\left( \dfrac{1/z_1}{1/z_1+ \ldots+ 1/z_d}, \cdots, \dfrac{1/z_d}{1/z_1+ \cdots+ 1/z_d}\right).
\end{equation}
Given $\omegaV$, \eqref{eq:pickands} implies $V(1/\omegaV)=A(\omegaV)$. 
We denote by $M_{\omegaV}^{\max}=\max_{i=1}^d \omega_i Z_i$ the max-projection of the random vector $\mathbf{Z}$. Then $M_{\omegaV}^{\max}$ is Fr\'{e}chet distributed with scale parameter $V(1/\omegaV)\leq 1$ reflecting the level of dependence in $\mathbf{Z}$ at an angle $\omegaV$: For $\mathbf{z}>0$, we get 
\begin{eqnarray}
\Pr \left( M_{\omegaV}^{\max} \leq z\right) %&=& \Pr\left( \omega_1 X^{*}_1 \leq z, \ldots, \omega_d X^{*}_d \leq z \right) %\notag \\
&=& \Pr\left( Z_1 \leq \dfrac{z}{ \omega_1}, \ldots, Z_d \leq \dfrac{z}{ \omega_d} \right) \notag \\
&=& \exp\left\lbrace -V\left( \dfrac{z}{ \omega_1}, \ldots, \dfrac{z}{ \omega_d}\right) \right\rbrace = \exp\left\lbrace -\dfrac{1}{z} V\left( \dfrac{1}{\omega_1},\ldots,\dfrac{1}{\omega_d}\right) \right\rbrace, \label{M_tilde_AD}
\end{eqnarray}
From \eqref{eq:pickands}, we conclude that the scale parameter is equal to $A_{\omegaV}:=A(\omegaV)$, and we remark that the corresponding min-projection $M^{\min}_{\omegaV}=1/M_{\omegaV}^{\max}$ follows an exponential distribution with rate $A_{\omegaV}$. As argued by \citet{ledford_1997}, a censored version of the multivariate max-stable distribution is a natural model for asymptotic dependent threshold exceedances. In view of convergence \eqref{eq:mrv}, this corresponds to applying the approximation $F(\mathbf{x})=\exp\{-V(\mathbf{x})\}\approx 1-V(\mathbf{x})$ for large values of $\|\mathbf{x}\|$ or equivalently, to replacing $1-F(t\bm z)$ by $-\log F(t\bm z)$ using the logarithmic series approximation $\log(1-\varepsilon)\approx -\varepsilon$ for small $\varepsilon>0$. 
We propose to consider projected data values $M^{\min\downarrow}_{\omegaV}=1/\max_{i=1}^d \omega_i X^F_i=\min_{i=1}^d X_i^{E\downarrow}/\omega_i$ using standard exponentially distributed $X_i^{E\downarrow}=1/ X^F_i$, where we use "$\downarrow$" to emphasize the tail inversion. As left-censoring the upper tail is equivalent to right-censoring the lower tail after inverting the tails, we model, below a small fixed threshold, the projected values by the $\mathrm{Exp}(A_{\omegaV})$ distribution, while we censor the values above the threshold. 

Our max-projection $M_{\omegaV}^{\max}$ is closely related to the max-projection $Y(\omegaV)$ used by \cite{Mhalla2017} for estimating max-stable dependence from maxima data where no censoring is applied. They define
\begin{equation}\label{eq:compproj}
Y(\omega_1,\omega_2) = \max \left[ \exp\{-1/(\omega_2 X_1)\}, \exp\{-1/(\omega_1 X_2)\}\right] 
\end{equation}
which corresponds to $\exp\{- \max(\omega_2 X_1,\omega_1 X_2)^{-1}\}=\exp\{-1/M_{(\omega_2,\omega_1)}^{\max}\}$. The link between the distribution functions of the two max-projections arises from the fact that $Z \sim \text{Exp}(\lambda)$ implies $\exp(-Z) \sim \text{Beta}(\lambda,1)$ for a random variable $Z$. The components of the direction $\omegaV$ in \eqref{eq:compproj} are inversed as we define the Pickands dependence function $A_{\omegaV}$ differently from \cite{Mhalla2017}.

%where $1-V(\mathbf{x})$ is (up to a rescaling) the distribution function of the multivariate Pareto limit model for threshold exceedances \citep{Rootzen.al.2017}. 

\subsection{Residual dependence in asymptotic independence}
\label{sec:ai}
In this section, we consider marginal transformations of the data vector $\mathbf{X}=(X_1,\ldots,X_d)$ to either standard Pareto margins $\mathbf{X}^P=(X_1^P,\ldots,X_d^P)$ or to unit exponential margins $\mathbf{X}^E=(X_1^E,\ldots,X_d^E)=\log(\mathbf{X}^P)$. We define the min-projection of $\mathbf{X}^E$ as $M_{\omegaV}^{\min}=\min_{i=1}^{d} X^E_i/\omega_i$ for a direction $\omegaV=(\omega_1,\ldots,\omega_d)$ in $S_{d}$. Based on the multivariate tail representation \eqref{AI_WT}, the function
\begin{equation}
f(x;\omegaV) = \Pr\left(X^P_1 > x^{\omega_1}, \ldots, X^P_d > x^{\omega_d}\right) = \Pr\left( M_{\omegaV}^{\min} > \log x \right), \quad \text{for } x \geq 1, \notag
\end{equation}
is regularly varying at infinity with index $\lambda_{\omegaV}$. Thus,
\begin{eqnarray}
\dfrac{f(tx;\omegaV)}{f(t;\omegaV)} &=& \dfrac{\Pr\left( M_{\omegaV}^{\min}> \log x + \log t\right) }{\Pr\left( M_{\omegaV}^{\min}>\log t\right) } \notag \\
&=& \Pr\left( M_{\omegaV}^{\min}> \log x + u \vert M_{\omegaV}^{\min}> u\right) \rightarrow x^{-\lambda_{\omegaV}}, \quad \text{as } u \rightarrow + \infty, \label{slow_var_AI}
\end{eqnarray}
where $u=\log t \rightarrow + \infty$ as $t \rightarrow + \infty$. Equivalently, the excesses of the structure variable $M_{\omegaV}^{\min}$ above a high threshold $u$ are exponentially distributed with rate $\lambda_{\omegaV}$ in the limit: Setting $\tilde{x}=\log x$ in \eqref{slow_var_AI} yields
\begin{equation}
\Pr\left( M_{\omegaV}^{\min} > \tilde{x} + u \vert M_{\omegaV}^{\min}> u \right) \rightarrow \exp\left\lbrace -\tilde{x} \lambda_{\omegaV}\right\rbrace, \quad \text{as } u \rightarrow + \infty. \label{M_tilde_AI}
\end{equation}
By using the tail structure \eqref{AI_WT} for characterizing asymptotic independence and appropriate marginal pretransformations, we can therefore model the positive excess $M_{\omegaV}^{\min} - u$ of the min-projection above a fixed high threshold $u$ through an $\mathrm{Exp}(\lambda_{\omegaV})$-distribution. 

A more specific yet still very flexible class of asymptotically independent processes satisfying \eqref{AI_WT} are the \emph{inverted max-stable processes} discussed in \cite{wadsworth_2012}. In unit Pareto margins, the limit relation \eqref{AI_WT} is exact for those models and can be written
\begin{equation}
\Pr\left( X^P_1 > x^{\omega_1}, \ldots, X^P_d > x^{\omega_d}\right) = x^{-A_{\omegaV}}, \quad x \geq 1, \label{tail_IEVD}
\end{equation}
with $A_{\omegaV}$ the Pickands dependence function of the associated max-stable process, which takes the role of the angular dependence function $\lambda_{\omegaV}$. If $\mathbf{X}^P$ is an inverted max-stable process, then the original max-stable process is recovered through the transformation $1/\log(\mathbf{X}^P)$. The slowly varying function $\ell(.;\omegaV)$ in \eqref{AI_WT} is equal to $1$ in this special case, such that
\begin{equation}\label{eq:expims}
M_{\omegaV}^{\min} \sim \text{Exp}\left\lbrace A_{\omegaV}\right\rbrace. 
\end{equation}
Therefore, if we assume an inverted max-stable dependence in the joint tail of $\mathbf{X}^P$, we can model data values of $M_{\omegaV}^{\min}$ above a fixed high threshold $u$ through the exponential distribution in \eqref{eq:expims} while censoring values below $u$. This is different from the general setup with arbitrary unknown $\ell(.;\omegaV)$ where only the excesses above the threshold are used, not the information contained in the censoring indicator $\mathbbm{1}_{M_{\omegaV}^{\min}>u}$.

\subsection{The case of Gaussian dependence} 
\begin{figure}[!h]
	\begin{center}
		\includegraphics[width=9cm,height=7cm]{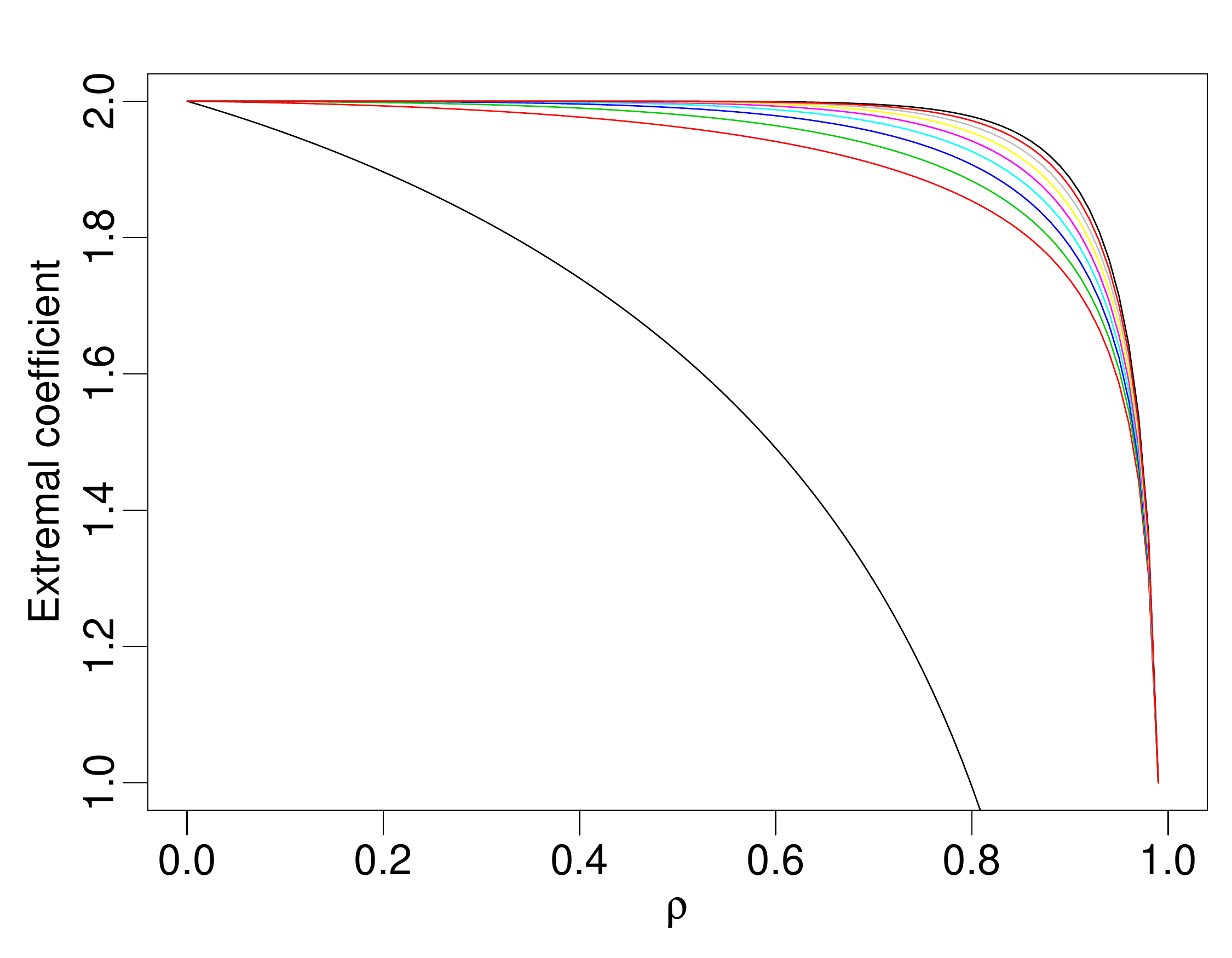}
		\caption{Sub-asymptotic extremal coefficient $\theta(r)$ of the Gaussian dependence as a function of the correlation $\rho$. Threshold levels are $r=-1/\log(1-10^{-q})$, with higher curves corresponding to higher levels}
		\label{theta_rho}
	\end{center}
\end{figure}

Suppose that $X_1^{F}$ and $X_2^{F}$ are unit Fr\'{e}chet distributed. On one hand, \cite{ledford_1996} showed that if $X_1^{F}$ and $X_2^{F}$ have bivariate normal dependence with correlation $\mathit{\rho<1}$, then
\begin{equation}
\Pr(X_1^{F}>r,X_2^{F}>r) \sim (1+\rho)^{3/2} (1-\rho)^{-1/2} (4\pi)^{-\rho/(1+\rho)} r^{-2/(1+\rho)} (\log r)^{-\rho/(1+\rho)}, \quad r \rightarrow \infty. \label{tail_Gaussian}
\end{equation}
On the other hand, if the distribution function of $\mathbf{X}^{F}=(X_1^{F},X_2^{F})$ is a bivariate extreme value distribution $F(x_1,x_2)=\exp\{-V(x_1,x_2)\}$ with exponent function $V$ and extremal coefficient $\theta=V(1,1)$, then
\begin{equation}
\Pr(X_1^{F}>r,X_2^{F}>r) \sim \left\lbrace 2-V(1,1)\right\rbrace r^{-1} + \left[ \left\lbrace V(1,1)\right\rbrace ^2/2-1\right] r^{-2}, \quad r \rightarrow \infty. \label{tail_EVD}
\end{equation}
By assuming asymptotic dependence for data from a Gaussian copula with $\rho<1$, we would fit an extreme value model at a sub-asymptotic level. Such model misspecification will result in biased extremal coefficient estimates smaller than $2$, the value for asymptotic independence, owing to residual dependence in the data at finite levels. We can approximately quantify this bias by equating \eqref{tail_Gaussian} and \eqref{tail_EVD}. The resulting \emph{subsasymptotic extremal coefficient} $\theta(r)$ is equal to
\begin{equation}
\theta(r) = r - \left\lbrace r^2 -4r+2+2(1+\rho)^{3/2}(1-\rho)^{-1/2} (4\pi)^{-\rho/(1+\rho)} r^{-2/(1+\rho)} (\log r)^{-\rho/(1+\rho)} \right\rbrace ^{1/2}, \label{theta_rho_relation}
\end{equation}
with $r$ the marginal threshold. This relationship between $r$, $\theta(r)$, and $\rho$ is displayed in Figure \ref{theta_rho} where the threshold $r$ is set to high quantiles of the unit Fr\'{e}chet distribution, $r=-1/\log(1-10^{-q})$ with $q=1,\ldots,10$.
For any $\rho<1$, the coefficient $\theta(r)$ tends to $2$ when $r$ tends to infinity. This is obtained from an application of the binomial series formula to approximate the right-hand side of \eqref{theta_rho_relation}. 

The extremal coefficient behaves as expected in the limit cases of perfect independence ($\rho=0$) and perfect dependence ($\rho=1$). By taking high thresholds such that residual dependence in the exceedance data vanishes, the extremal coefficient is close to $2$ unless the correlation $\rho \approx 1$. Detection of asymptotic independence in the Gaussian dependence case with finite sample size was studied in \cite{bucher2011} where the authors proposed new estimators of the Pickands dependence function.

\subsection{An illustration on data}
In Sections~\ref{sec:ad} and \ref{sec:ai}, we developed asymptotically justified univariate exponential models for both classes of asymptotic dependence and independence using projections. 
Equation \eqref{AI_WT} characterizes the joint tail behavior of random vectors with 
\begin{eqnarray}
\lambda_{\omegaV} &\in& ]\max_{1\leq i \leq d} \omega_i ,1], \quad \text{or } \lambda_{\omegaV} = \max_{1\leq i \leq d} \omega_i, \ \ell(x;\omegaV) \rightarrow 0 \quad \text{(asymptotic independence),} \notag \\
\lambda_{\omegaV} &= & \max_{1\leq i \leq d} \omega_i \quad \text{and } \ell(x;\omegaV) \not\rightarrow 0 \quad \text{(asymptotic dependence),} \notag
\end{eqnarray}
where $\ell(x;\omegaV)$ has a well-defined positive limit in the asymptotic dependence case if the regular variation property \eqref{eq:mrv} holds. If we look at data under the assumption of asymptotic dependence, then we calculate the projections $M_{\omegaV}^{\min\downarrow}$ to estimate the values of the Pickands dependence function. If we start with the assumption of asymptotic independence, then we calculate $M_{\omegaV}^{\min}$ to estimate the values of the angular dependence function. A max-stable model characterized by the Pickands function $A$ provides an appealing link between the two dependence classes by the possibility of modeling asymptotic independence with the corresponding inverted max-stable model whose function $\ell(\cdot;\omegaV)\equiv 1$ is known a priori and whose angular dependence function is $A$. 
Notice that the link between the two projections is as follows, using probability integral transformations: 
$$M_{\omegaV}^{\min\downarrow}=-\frac{\log\left\lbrace 1-\exp\left(-M_{\omegaV}^{\min}\right)\right\rbrace }{A(\boldsymbol{\omega})}.$$

\begin{figure}
 \centering
  \centering 
\subfloat{\includegraphics[width=0.33\linewidth]{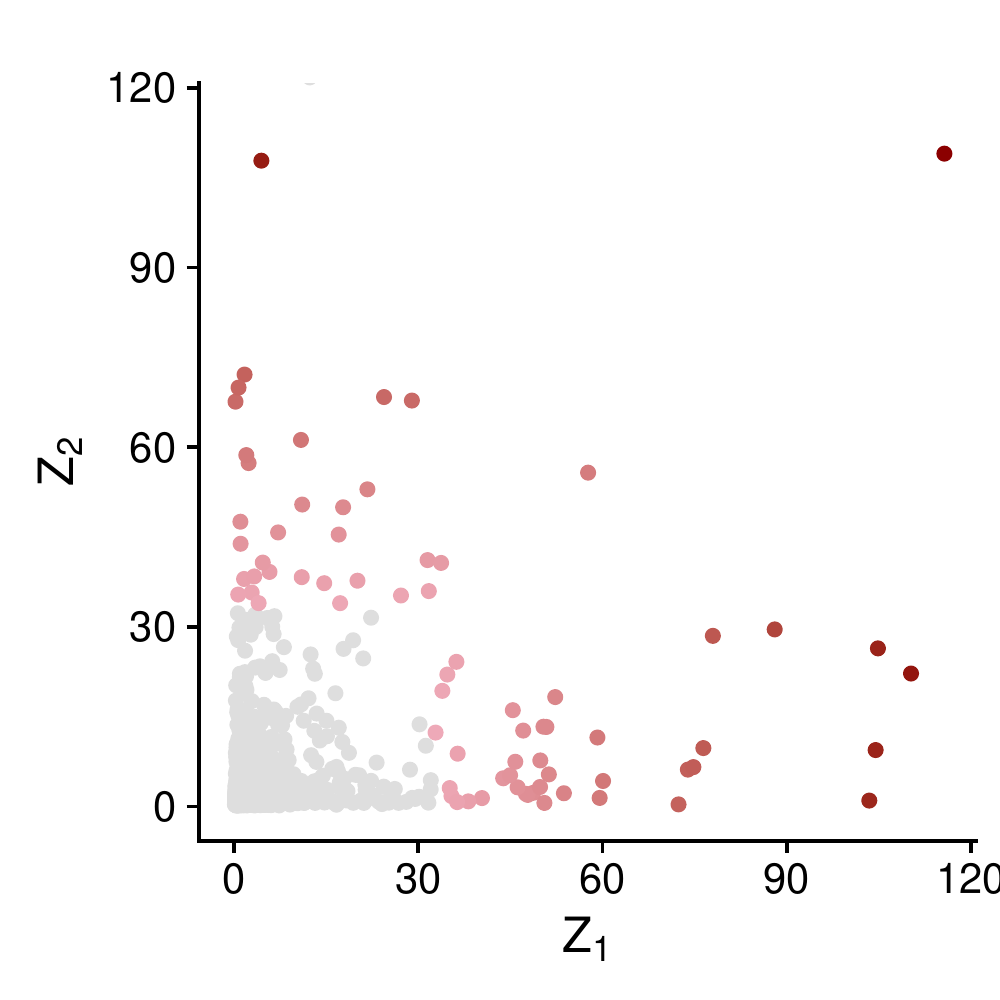}}
\subfloat{\includegraphics[width=0.33\linewidth]{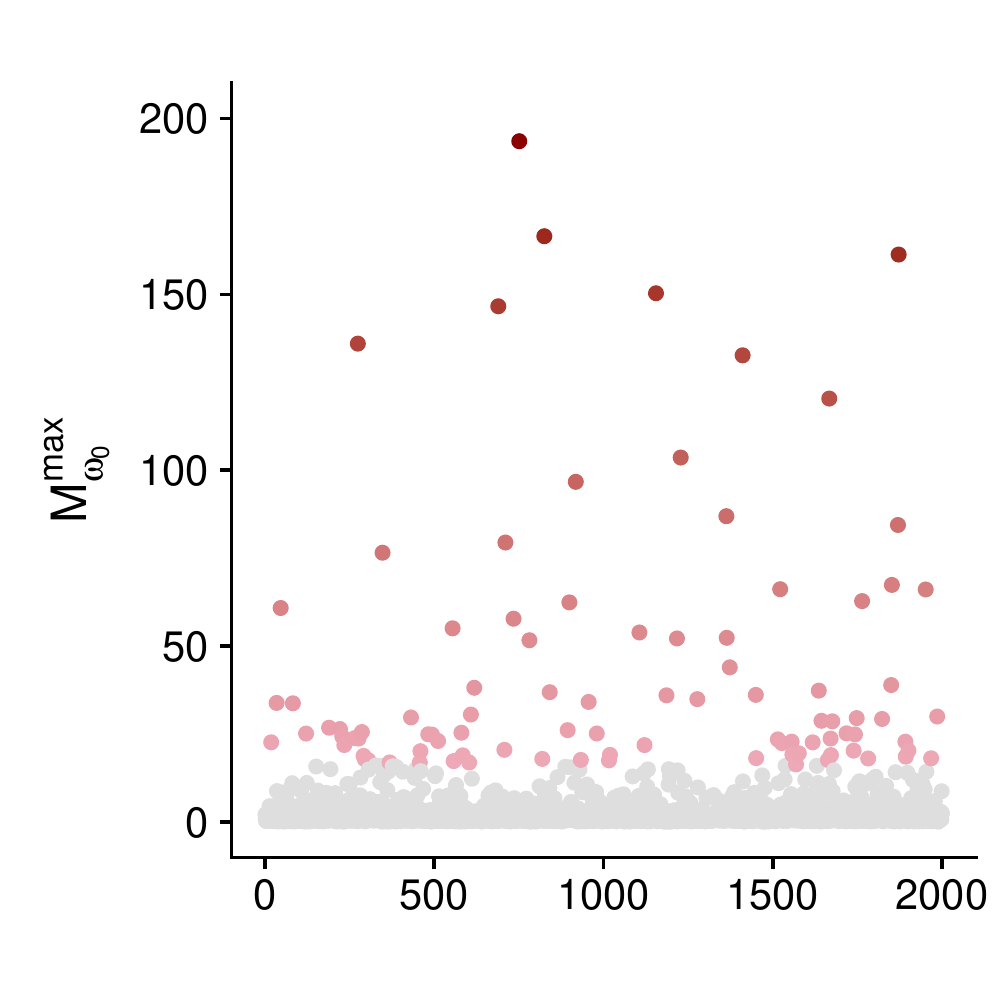}}
\subfloat{\includegraphics[width=0.33\linewidth]{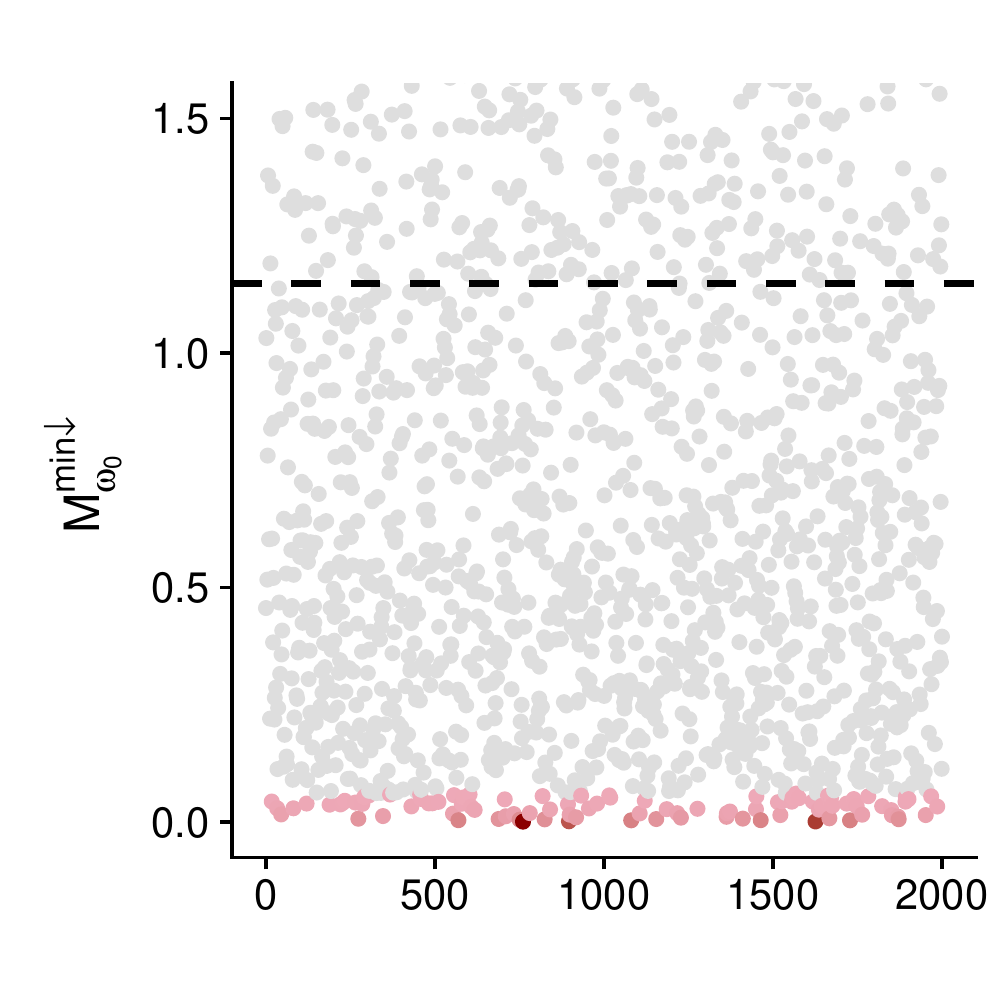}}\\
\subfloat{\includegraphics[width=0.33\linewidth]{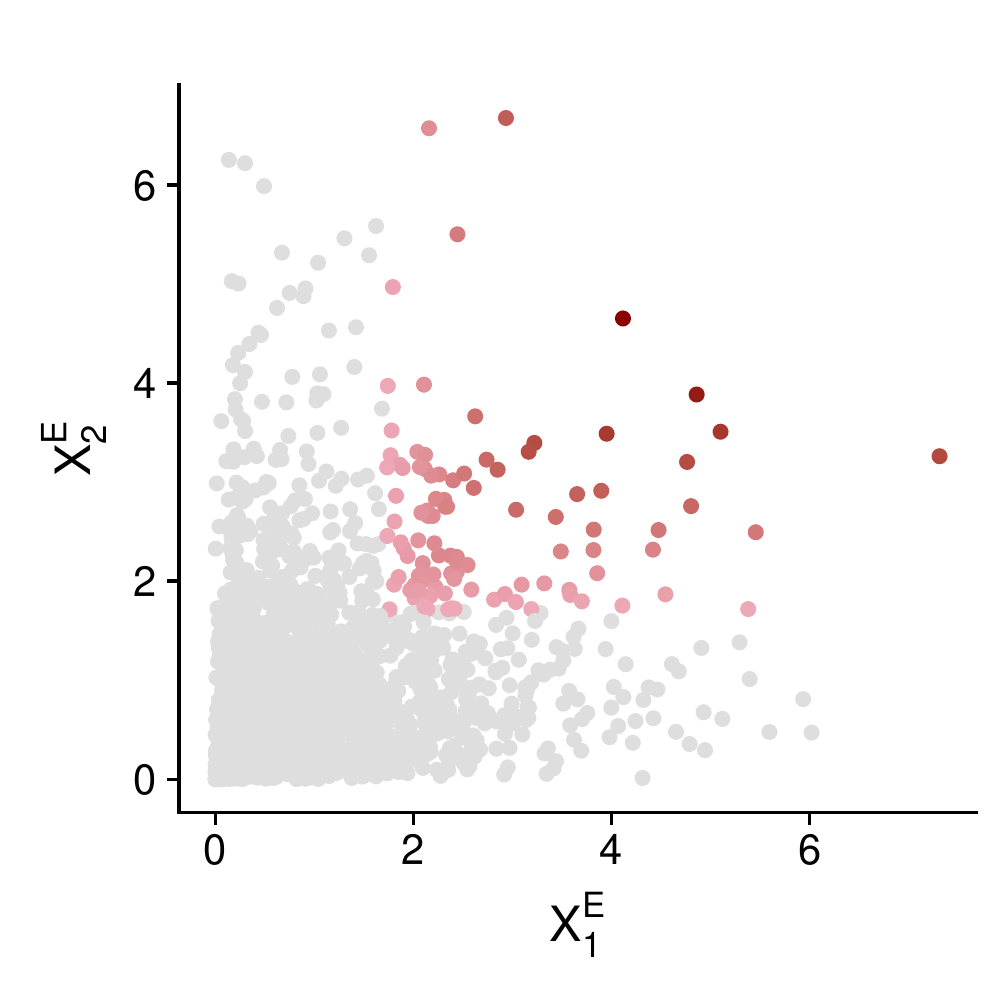}}
\subfloat{\includegraphics[width=0.33\linewidth]{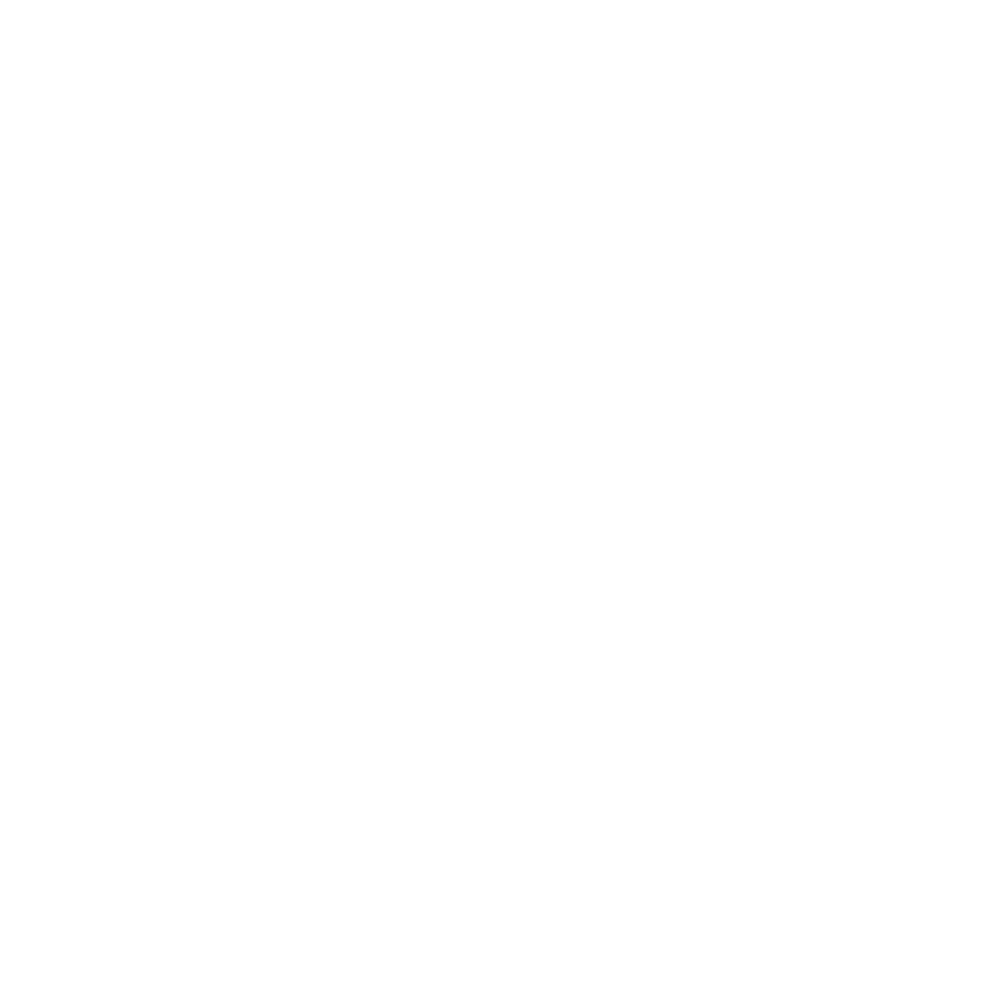}} \hspace*{0.2cm}
\subfloat{\includegraphics[width=0.33\linewidth]{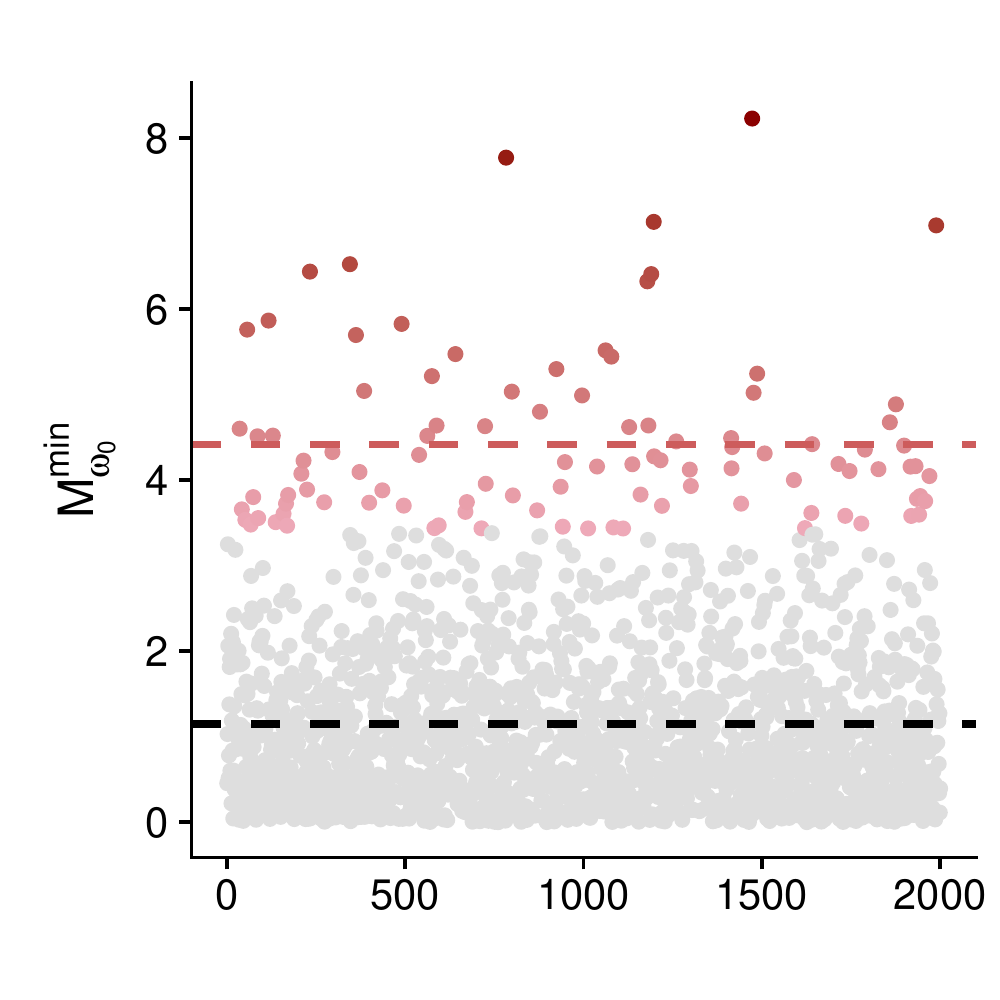}}
 \caption{Realizations of a bivariate logistic extreme value distribution ($n=2000$) with parameter $0.8$ on the unit Fr\'{e}chet scale (top left) and the standard exponential scale (bottom left), its corresponding max-projection (middle), and its corresponding min-projection (right). The \textit{red} points correspond to the deficits of $M_{\omegaV_0}^{\min\downarrow}$ below the $5\%$ quantile (top) and the exceedances of $M_{\omegaV_0}^{\min}$ above the $95\%$ quantile (bottom). The dashed \textit{black} lines indicate the mean of the random variables $M_{\omegaV_0}^{\min\downarrow}$ and $M_{\omegaV_0}^{\min}$, i.e., the inverse of $A_{\omegaV_0}$ for the logistic model. The dashed \textit{red} line corresponds to the mean of the excesses of $M_{\omegaV_0}^{\min}$ \label{excess_max_stable}}
\end{figure}

Figure \ref{excess_max_stable} illustrates these relationships on bivariate data simulated according to an extreme value distribution, where various marginal scales and projections at $\omegaV_0=(1/2,1/2)$ are considered. Empirical thresholds at the $95\%$ and $5\%$ levels are used to censor the upper and lower tails, respectively. 

\section{Inference and regression modeling of dependence}
\label{sec:inference}
The theory above shows that the projection techniques are appropriate for modeling asymptotic dependence and asymptotic independence in threshold excesses. In either case, we aim to develop a dependence model for the upper joint tail of the data distribution. With asymptotic independence, we transform the data margins to the exponential distribution, calculate the min-projection, and model its excesses above a high threshold through an exponential distribution, whose rate is given by the angular dependence function evaluated at the projection angle. More specifically, the latter corresponds to the Pickands dependence function if we utilize an inverted max-stable model. With asymptotic dependence, we transform the data margins to the unit Fr\'echet distribution, calculate the max-projection, invert the latter, and model the resulting deficits below a small fixed threshold through an upper-censored exponential distribution, whose rate is given by the Pickands dependence function evaluated at the projection angle.

By using min- and max-projections, inference is based on a univariate variable, which frees us from handling multivariate censoring schemes in likelihood-based approaches and the resulting computational burden. A difference between the two dependence regimes arises in the construction of the projection. On one hand, the max-projection in the case of asymptotic dependence retains only the highest value, which makes sense since the asymptotic theory implies that the limit model gives a good approximation for the components that are large. On the other hand, the min-projection in the case of asymptotic independence retains the smallest value, which contains crucial information on the faster tail decay rate along various directions $\omegaV$, while the max-projection would converge to a unit exponential limit carrying no useful information in this case. 

In either case, inference for the dependence structure is a two-step procedure based on a univariate structure variable. In the first step, the margins are transformed to the unit Fr\'{e}chet scale and then inverted to the unit exponential scale (asymptotic dependence) or directly to the unit exponential scale (asymptotic independence), and we then calculate the min-projection. The second step consists of fitting the appropriate exponential model \eqref{M_tilde_AD} or \eqref{M_tilde_AI} to $M^{\min\downarrow}_{\omegaV}$ and $M^{\min}_{\omegaV}$, respectively. In the following, we detail the second step of the inference procedure with a view towards estimating the influence of a set of covariates $\mathbf{y} \in \mathbb{R}^q$. 

\subsection{Inference for the case of asymptotic dependence}
Given $n$ observations $\{\mathbf{x}^{F}_j\}_{j=1}^n$ of the random vector $\mathbf{X}^{F}=(X^{F}_1,\ldots,X^{F}_d)$ with unit Fr\'{e}chet margins, we suppose that $\mathbf{X}^{F}$ is in the max-domain of attraction of an extreme value distribution with Pickands dependence function $A$. We fix a direction $\omegaV=(\omega_1,\ldots,\omega_d) \in S_d$ and calculate the observed structure variables $m^{\min\downarrow}_{\omegaV,j}=\min_{i=1}^d 1/ (\omega_i x_{j,i}^{F})$, $j=1,\ldots,n$. To put focus on the dependence of extremes in the second step of the inference procedure, we censor observations $m^{\min\downarrow}_{\omegaV,j}$ that are above a low threshold $u>0$, for instance, chosen as the empirical $5\%$ quantile of $m^{\min\downarrow}_{\omegaV,j}$. The likelihood function is $L\left( A_{\omegaV};m^{\min\downarrow}_{\omegaV,1},\ldots,m^{\min\downarrow}_{\omegaV,n} \right)  = \prod_{j=1}^{n} L_j \left( A_{\omegaV}\right) $ with contributions
$$L_j \left( A_{\omegaV}\right) = \left\{
  \begin{array}{ll}
    A_{\omegaV} \exp\left( -A_{\omegaV}m^{\min\downarrow}_{\omegaV,j} \right) & \mbox{if } m^{\min\downarrow}_{\omegaV,j} < u, \\
    \exp \left( -A_{\omegaV} u \right) & \mbox{if } m^{\min\downarrow}_{\omegaV,j} \geq u.
  \end{array}
\right.$$
When a set of covariates $\mathbf{y}\in \IR^q$ is available, we propose using a generalized additive model (GAM) structure \citep{wood_Book_2017} to model the dependence in the extremes, similar to the approach of \citet{Mhalla2017}, who estimated the Pickands dependence function based on the block maxima approach. The dependence of $\mathbf{x}^{F}$ on $\bf{y}$ is assumed to be at the extremal dependence level, i.e., the covariates solely influence the Pickands dependence function $A_{\omegaV} \equiv A_{\omegaV}(\mathbf{y})$. This assumption implies no loss of generality as the marginal inference is performed in a separate step. A very general model for $A_{\omegaV}(\mathbf{y})$ arises from supposing the semi-parametric form
\begin{eqnarray}
A_{\omegaV}(\mathbf{y};\mathbf{\Lambda}) & = & h^{-1}\left\{\mathbf{u}^{T}\greekvec{\beta} + \sum_{k=1}^K h_{k}(t_k)\right\},
\label{eq:condsemipara}
\end{eqnarray}
where $h$ is a link function and $\mathbf{u}\in \IR^s$ and $(t_1,\ldots,t_K)$ are subvectors of $\mathbf{y}$, or products of covariates if interactions between some covariates are considered. The column vector $\greekvec{\beta} \in \IR^s$ gathers linear coefficients whereas $h_{k}: \mathbb{H}_{k} \rightarrow \IR $ are smooth functions supported on closed intervals $\mathbb{H}_{k} \subset \IR$ and admitting a finite quadratic penalty representation \citep{greensilverman2000}. The column vector $\mathbf{\Lambda}$ gathers all parameters to be estimated in the model, i.e., the vector $\greekvec{\beta}$ and the linear basis coefficients of each of the smooth functions $h_{k}$. Based on a sample $\{\mathbf{x}^{F}_j, \mathbf{y}_j \}_{j=1}^n$, we estimate the GAM \eqref{eq:condsemipara} by maximizing the penalized log-likelihood
\begin{align}
\ell(\mathbf{\Lambda}, \boldsymbol{\gamma}) &= \ell(\mathbf{\Lambda}) - \frac{1}{2} \sum_{k=1}^K \gamma_{k}\int_{\mathbb{H}_{k}} h_{k}^{''}(t_k)^2\,\mathrm{d}t_k, 
\label{eq:lp}
\end{align}
with $\ell(\mathbf{\Lambda}) = \sum^n_{j=1}\ell_j(m^{\min\downarrow}_{\omegaV,j},\mathbf{y}_j,\mathbf{\Lambda})$, 

$$\ell_j(m^{\min\downarrow}_{\omegaV,j},\mathbf{y}_j,\mathbf{\Lambda})=\left\{
  \begin{array}{ll}
    \log{A_{\omegaV}(\mathbf{y}_j)} -A_{\omegaV}(\mathbf{y}_j)m^{\min\downarrow}_{\omegaV,j} & \mbox{if } m^{\min\downarrow}_{\omegaV,j} < u, \\
    -A_{\omegaV}(\mathbf{y}_j) u & \mbox{if } m^{\min\downarrow}_{\omegaV,j} \geq u.
  \end{array}
\right.$$ The integrals in \eqref{eq:lp} are componentwise roughness penalties with smoothing parameters $\mathbf{\gamma}=(\gamma_{1},\ldots,\gamma_{K})$ that balance between the smoothness of the model and its goodness of fit. Higher values of $\gamma_{k}$ yield smoother fitted curves. The related effective degrees of freedom of each smooth function $h_k$ are defined as ${\rm trace}\left(  I + \gamma_k S_k\right)$, where $S_k$ is the positive definite penalty matrix associated to the basis representation of $h_k$ \citep[Chapter 5]{wood_Book_2017}. The maximization of the penalized log-likelihood \eqref{eq:lp} is performed based on an outer-iteration procedure. At each iteration, $\mathbf{\Lambda}$ and $\boldsymbol{\gamma}$ are estimated separately by penalized iteratively re-weighted least squares (PIRLS) and a prediction error method (Generalized Cross Validation), respectively; see \cite{wood_Book_2017} for a detailed description of the available methods for GAM fitting. \\
The penalized maximum log-likelihood estimator $\widehat{\mathbf{\Lambda}}_n $ then defines the estimate $\hat{A}_{\omegaV}(\mathbf{y})= h^{-1}\left\{\mathbf{u}^{T}\hat{\greekvec{\beta}} + \sum_{k=1}^K \hat{h}_{k}(t_k)\right\}$ of the Pickands dependence function evaluated at $\omegaV$. A vast amount of literature on GAM-related theory is available and includes \cite{Wood_2004,Wood_2006,Marra2011}, among others.
%\label{eq:pmle}

%with $\ell(\vtheta) = \sum^n_{i=1}\ell_0(u_{i1},u_{i2}, \mathbf{w}_{i};\vtheta)$, $\vgamma=(\gamma_1, \ldots, \gamma_K)^{\top}$ and $\gamma_k \in \IR_{+} \cup \left\lbrace 0 \right\rbrace$ for all $k$. The integral terms are roughness penalties on each component and $\vgamma$ is a vector of smoothing parameters. The penalized maximum log-likelihood estimator is defined as
%\begin{align*} \widehat{ A_{\omegaV}}_n = \underset{\displaystyle \vtheta \in \Theta}{\mbox{argmax }} \ell(\vtheta, \vgamma).
%\label{eq:pmle}
%\end{align*}

\subsection{Inference for the case of asymptotic independence}
Given $n$ observations $\{\mathbf{x}^P_j\}_{j=1}^n$ of the random vector $\mathbf{X}^P=(X_1^P,\ldots,X_d^P)$ with standard Pareto margins, we suppose that $\mathbf{X}^P$ has an asymptotic independent tail structure as in \eqref{AI_WT}. 
We fix a direction $\omegaV=(\omega_1,\ldots,\omega_d) \in S_{d}$ and calculate the observed structure variables $$m_{\omegaV,j}^{\min}=\min_{i=1}^{d}\log(x^P_{j,i})/\omega_i, \quad j=1,\ldots,n.$$ We fix a high threshold $u$, for instance, chosen as the the empirical $95\%$ quantile of $m_{\omegaV,j}^{\min}$, and extract the sample of positive excesses $\tilde{m}_{\omegaV,j_e}^{\min}=m_{\omegaV,{j_e}}^{\min}-u>0$, $e=1,\ldots,E_u$ with a positive number of excesses $E_u>0$. 
Then, we maximize the likelihood composed of contributions
\begin{equation}
L_{j_e}\left\lbrace \lambda_{\omegaV}\right\rbrace = \lambda_{\omegaV} \exp\left\lbrace -\lambda_{\omegaV} \left( \tilde{m}^{\min}_{\omegaV,j_e}\right) \right\rbrace, \quad e=1,\ldots E_u.
\label{lik.aind}
\end{equation}
Given covariate vectors $\mathbf{y}_j$, we proceed as for the asymptotic dependence case by proposing a GAM \eqref{eq:condsemipara} for $\lambda_{\omegaV} \equiv \lambda_{\omegaV}(\mathbf{y})$, i.e.,
\begin{equation}
\lambda_{\omegaV}(\mathbf{y};\mathbf{\Lambda}) = h^{-1}\left\{\mathbf{u}^{T}\greekvec{\beta} + \sum_{k=1}^K h_{k}(t_k)\right\},
\notag
\end{equation}
which is fitted by maximizing a penalized version of the likelihood \eqref{lik.aind} similarly to \eqref{eq:lp} and results in the estimate $\hat{\lambda}_{\omegaV}(\mathbf{y})$ of the tail dependence function evaluated at $\omegaV$.

\section{Simulation study}
\label{sec:simu}
We study the properties of the estimators $\hat{A}_{\omegaV}(\mathbf{y})$ and $\hat{\lambda}_{\omegaV}(\mathbf{y})$ when they are related to a covariate vector $\mathbf{y} \in \mathbb{R}^q$ using the link function $h(x)=\log\{(x-1/2)/(1-x)\}$, a modification of the $\mathrm{logit}$ link resulting in values within $(0.5,1)$. We focus on the estimation of these two dependence functions in the bivariate case at $\omegaV_0=(1/2,1/2)$, which yields estimates of the covariate-dependent coefficients of extremal dependence $\hat{\theta}(\mathbf{y})=2\hat{A}_{\omegaV_0}(\mathbf{y})$ and tail dependence $\hat{\eta}(\mathbf{y})=1/\{2\lambda_{\omegaV_0}(\mathbf{y})\}$. Realistic sample sizes are chosen, similar to those in the subsequent application. 
\subsection{Case of asymptotic dependence}
Suppose that $\mathbf{X}^{F}$ is an asymptotically dependent random vector with unit Fr\'{e}chet margins. We first consider the max-domain of attraction (MDA) setting where our model represents an asymptotic approximation of the exact tail behavior in the data. 
%We distinguish between the case where the copula $C$ is in the max-domain of attraction (MDA) of an extreme value (EV) copula without being max-stable itself, and the case where it is itself an EV copula. Our model is exact for the latter while it represents an asymptotic approximation of the former. 
\subsubsection{Data distribution in the maximum domain of attraction \label{4.1.1}}
We focus on the bivariate case with $\mathbf{X}^{F}=(X^{F}_1,X^{F}_2)$ a random vector with unit Fr\' {e}chet margins and an Archimedean copula $C^{\star}$ \cite[Chapter 4]{nelsen2006} with generator $\varphi(t)=(1/t-1)^{1/\alpha}$ for some $\alpha \in (0,1)$, i.e., the distribution function of $\mathbf{X}^{F}$ is 

\begin{equation}
F(x_1,x_2) = \varphi\left[ \varphi^{-1}\left\lbrace \exp\left( -\dfrac{1}{x_1}\right) \right\rbrace + \varphi^{-1}\left\lbrace \exp\left( -\dfrac{1}{x_2}\right) \right\rbrace\right] , \quad x_1,x_2>0. \label{archimedean_F}
\end{equation}
This distribution is in the max-domain of attraction of the logistic bivariate extreme value distribution \citep{logistic} with parameter $\alpha$ \citep{fougeres_2004}, and can be simulated using the algorithm from \citet[example 4.15, p. 144]{nelsen2006}. We estimate the covariate-dependent extremal coefficient using deficits of $M^{\min\downarrow}_{\omegaV_0}$ by setting $\omegaV_0=(1/2,1/2)$ and fixing different threshold levels. 

%\begin{itemize}
%\item Generate two independent random variables $s$ and $t$ uniformly distributed on $(0,1)$. 
%\item Set $w=K_{C^{*}}^{-1}(t)$, where $K_{C^{*}}(x)= x- \dfrac{\varphi(x)}{\varphi^{'}(x)}$.
%\item Set $u=\varphi^{-1}\lbrace s \varphi(w)\rbrace$ and $v= \varphi^{-1} \lbrace (1-s) \varphi(w)\rbrace$.
%\item The desired pair is $(u,v)$.
%\end{itemize}
%In Algorithm 1, we have that
%\begin{eqnarray}
%\varphi(t) &=& \left( \dfrac{1}{t}-1\right) ^{1/\alpha}, \quad \alpha \in (0,1), \quad t \in (0,1], \notag \\
%\varphi^{-1}(x) &=& \dfrac{1}{x^{\alpha}+1}, \notag \\
%K_{C^{*}}(t) &=& t + \alpha t^2\left( \dfrac{1}{t}-1\right), \quad \text{and} \notag \\
%K_{C^{*}}^{-1}(x) &=& \dfrac{1+\alpha-\sqrt{(1+\alpha)^2-4\alpha x}}{2\alpha}. \notag
%\end{eqnarray}
Equation \eqref{M_tilde_AD} holds when at least one variable $X_i^{F}$ exceeds some high threshold, and the limiting dependence structure in the tails of $\mathbf{X}^{F}$ is equal to that of a bivariate logistic extreme value distribution \citep{Coles_1991}. We empirically study the bias of the covariate-dependent extremal coefficient estimator resulting from the application of the limiting extreme value model. 

\begin{figure}[!h]
\begin{center}
\includegraphics[width=9cm,height=7cm]{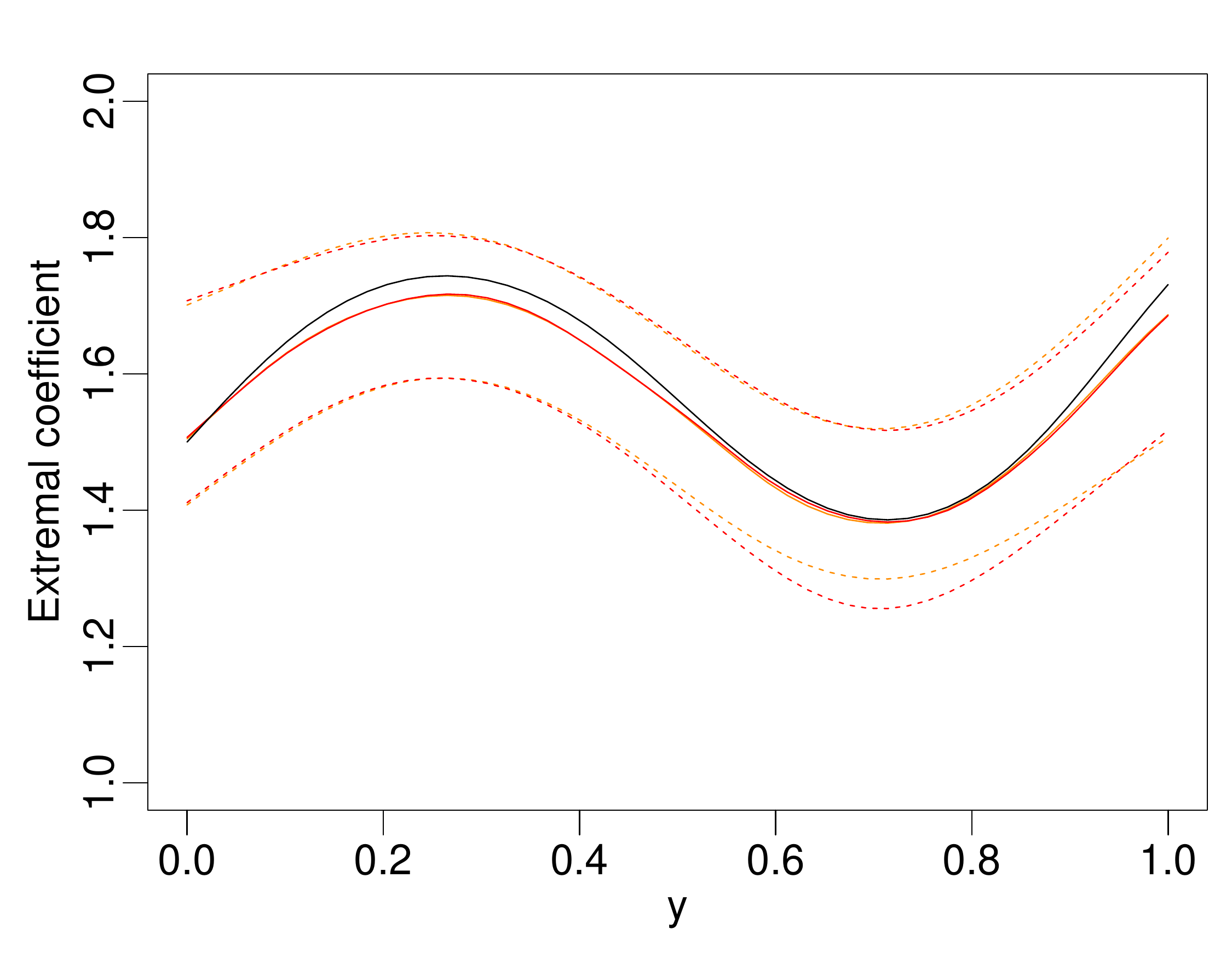}
\caption{Estimation in the max-domain of attraction setting. Estimates (solid) and $95\%$ bootstrap confidence intervals (dashed) of the extremal coefficient $\theta(y)$ in the first and second levels of $I$ in \textit{orange} and \textit{red} lines, respectively. The true values as given by \eqref{A_MDA} are in solid \textit{black} lines}
\label{MDA_5_levels}
\end{center}
\end{figure}

\begin{figure}[!h]
\begin{center}
\includegraphics[width=12cm,height=7cm]{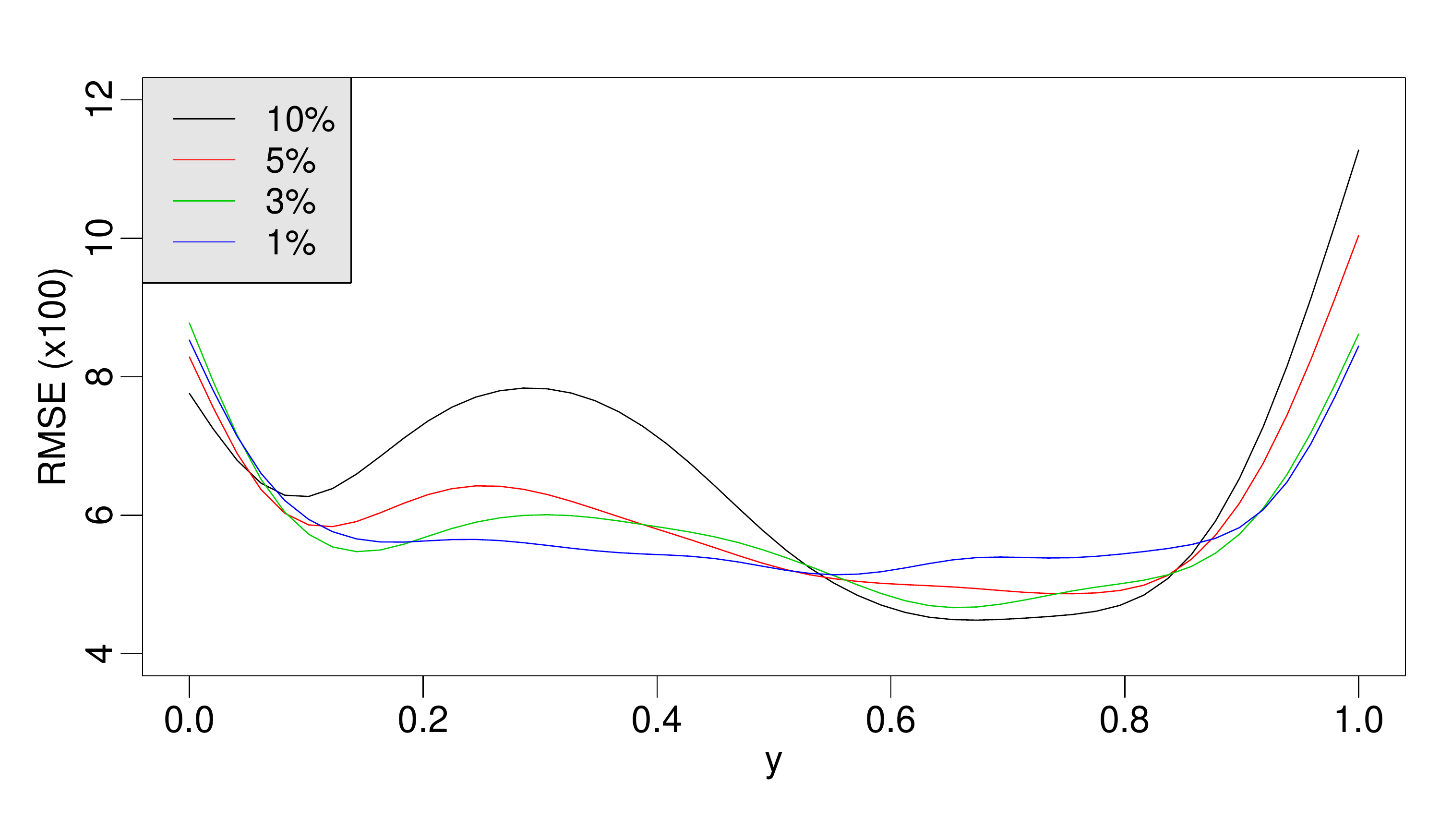}
\caption{Estimation in the max-domain of attraction setting, with the root mean squared error of the estimates of $\theta(y)$ for $y \in [0,1]$ based on $500$ samples and obtained at different threshold levels}
\label{RMSE_MDA}
\end{center}
\end{figure}

We simulate from a covariate-dependent model \eqref{archimedean_F} with the following generator
\begin{equation}
\varphi(t;y) = \left( \dfrac{1}{t}-1\right) ^{1/\alpha(y)}, \notag
\end{equation}
where $y \in [0,1]$ is observed at $50$ equally spaced values and 
\begin{equation}
\alpha(y) = \log\left[ 1+ \dfrac{\exp\left\lbrace \sin(2\pi y) + y^2\right\rbrace}{1+\exp\left\lbrace \sin(2\pi y) + y^2\right\rbrace}\right] /\log(2) \in (0,1). \notag
\end{equation}
Thus, the covariate-dependent extremal coefficient is
\begin{equation}
\theta(y)=2A_{\omegaV_0}(y)=1+\dfrac{\exp\left\lbrace \sin(2\pi y) + y^2\right\rbrace}{1+\exp\left\lbrace \sin(2\pi y) + y^2\right\rbrace}. \label{A_MDA}
\end{equation}
The sample size is chosen between $50000$ for a threshold at the $10\%$ level and $500000$ at the $1\%$ level, such that an average of $100$ threshold deficits of $M^{\min\downarrow}_{\omegaV_0}$ arises for a fixed value of $y$. Additionally, we empirically quantify the performance of our estimator in an overparametrized setting where one of the covariates has no influence on the response by considering a dummy categorical covariate $I$ with two levels. In the true model \eqref{A_MDA}, the Pickands dependence function does not depend on $I$. We simulate the observed values of $I$ at random and include level-dependent predictors in the GAM structure for $A_{\omegaV_0}(y)$; thus, the estimated model is
\begin{equation}
h\{\hat{A}_{\omegaV_0}(y,I)\}= \hat{a}_1 + \hat{s}_{1}(y) \mathbbm{1}_{\lbrace I=``\text{Level 1}"\rbrace} + \left\lbrace \hat{a}_2 + \hat{s}_2(y)\right\rbrace \mathbbm{1}_{\lbrace I=``\text{Level 2}"\rbrace}, \notag
\end{equation}
where $\hat{a}_i$, $i=1,2$ are the estimated intercepts and $\hat{s}_1$, and $\hat{s}_2$ are the estimated smooth curves describing the dependence of the Pickands function at $\omegaV_0$ on $y$ and at each level of $I$. The Monte Carlo procedure is based on $500$ repetitions, and $95\%$ percentile confidence intervals are constructed. Figure \ref{MDA_5_levels} displays the pointwise mean estimates of $\theta(y,I)$ in both levels of $I$ with a threshold set at the $5\%$ level. 
As expected, the inclusion of $I$ in the model does not prevent it from recovering the true dependence structure of the extremal coefficient on $y$ in both levels. Moreover, when considering the point estimates (i.e., the single-run experiments), the results indicate that the variable $I$ is not statistically significant in the fitted GAM. Next, we remove this artificial covariate from the model and compare the root mean squared error (RMSE) of the resulting extremal coefficient fits obtained for different threshold levels. The RMSE over the $500$ samples is defined as
\begin{equation}
\text{RMSE}(y)=\left[ \sum_{r=1}^{500} \{\hat{\theta}^{r}(y)-\theta(y)\}^2/500\right] ^{1/2}, \notag
\end{equation}
where $\hat{\theta}^{r}(y)$ is the covariate-dependent extremal coefficient estimate obtained from the $r$th sample. Figure \ref{RMSE_MDA} shows the RMSE of $\hat{\theta}(y)$ for different threshold levels. The RMSE tends to decrease when we take lower threshold levels. Similar shapes of the RMSE function are observed for threshold levels below $10\%$, with a noticeable decrease in the RMSE for values of $y$ corresponding to weak extremal dependence, i.e., for values of $y$ between $0.2$ and $0.4$. 

\subsubsection{Max-stable data distribution \label{4.1.2}}
\begin{figure}[!h]
\begin{center}
\includegraphics[width=12cm,height=7cm]{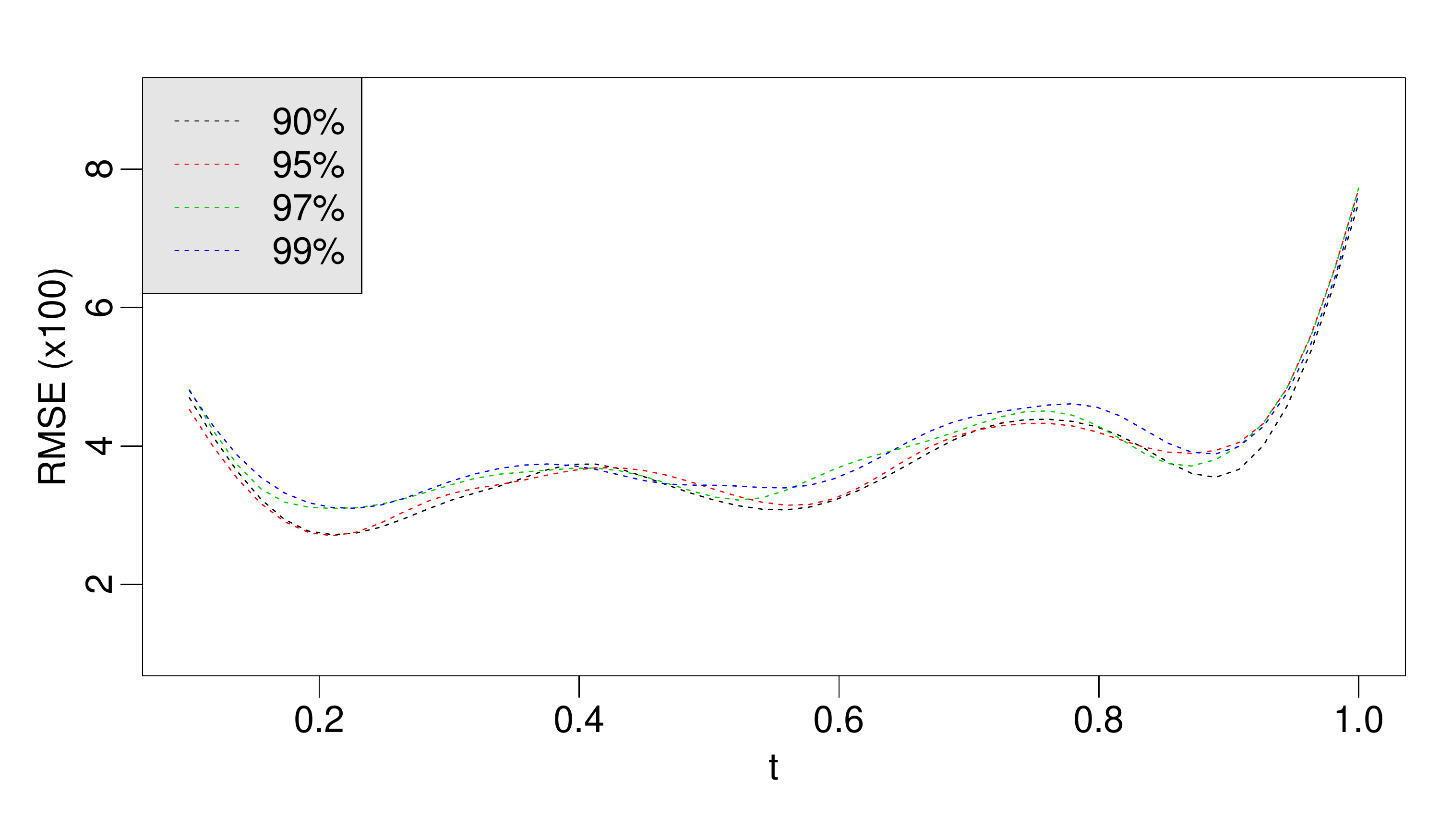}
\caption{Root mean squared error of the estimates of $\theta(t)$ for $t \in [0.1,1]$ with respect to different threshold levels, for $500$ simulations}
\label{gauss_bevd_AD}
\end{center}
\end{figure}

We now simulate bivariate samples with max-stable dependence where our asymptotic dependence model class contains the exact model. The observations come from the logistic extreme value copula with unit Fr\'{e}chet margins and dependence parameter $\alpha(t)=t-0.05$, $t \in [0.1,1]$. 
The covariate-dependent extremal coefficient $\theta(t)$ is $2^{\alpha(t)}$. Figure \ref{gauss_bevd_AD} displays the RMSE of the extremal coefficient estimates with respect to $t$ for the $10\%$, $5\%$, $3\%$, and $1\%$ threshold levels. The RMSE is mostly unaffected by the threshold level as our modeling assumption \eqref{M_tilde_AD} holds exactly, as opposed to the sub-asymptotic setting in Section~\ref{4.1.1}. Therefore, we observe a lower RMSE (see Figure~\ref{RMSE_MDA}), as there is no estimation bias resulting from penultimate modeling. 

%Regardless of the strength of the asymptotic dependence, the bias tends to be slightly larger when taking smaller threshold levels for the structure variable $M^{\min\downarrow}_{\omegaV_0}=\min(2/ X_1^{F},2/ X_2^{F})$, or equivalently higher threshold levels for $M^{\max}_{\omegaV_0}=\max(X^{F}_1/2,X^{F}_2/2)$. This might be due to the extreme region set on which we assume the asymptotic theory of multivariate extremes to be valid. As a matter of fact, in \eqref{M_tilde_AD}, we assume that the limit extreme value model is valid for large values of the structure variable $M^{\max}_{\omegaV}$, which is equivalent to assuming the validity of this model if at least one of the components of the random vector $\mathbf{X}^{F}$ is extreme. \cite{Huser_2014} showed in a comparative study of different likelihood threshold estimators for multivariate extremes that the bias resulting from the threshold estimation of the dependence parameter of the logistic model is larger when considering the exceedance set where at least one component is large, compared to the threshold estimator based on censoring the observations with at least one non-extreme component or equivalently applying the limit extreme value model when all the components are extreme\footnote{is it true?}. Moreover, the authors found that standard errors increase by increasing either the logistic dependence parameter or the threshold level. Their findings are consistent with the observed behavior of the standard deviation curves in Figure \ref{gauss_bevd_AD}.

\subsection{Case of asymptotic independence\label{4.2}}
We consider two models for an asymptotically independent random vector $\mathbf{X}^{E}=(X^{E}_1,X^{E}_2)$ with standard exponential margins. The dependence in the tails of $\mathbf{X}^{E}$ depends on the covariates $I$ and $t$, where $I$ is a categorical covariate with two levels $1$ and $2$. Bivariate Gaussian dependence with correlation $\rho(t)$ is observed in the first level of $I$, and an inverted logistic extreme value dependence with dependence parameter $\alpha(t)$ is observed in the second level of $I$. The dependence on covariates is as follows: 
\begin{eqnarray}
\rho(t) &=& t, \label{rho_t_I} \\
\alpha(t) &=& t-0.05, \label{alpha_t_I} \\
t &\in& [0.1,1]. \notag
\end{eqnarray}
Then, 
$$
\eta(t,I) = 1/\{2\lambda_{\omegaV_0}(t,I)\}= \left\{
\begin{array}{ll}
\{1+\rho(t)\}/2 & \mbox{if $I=``\text{Level 1}"$,} \\
1/\{2A_{\omegaV_0}(t)\}=2^{-\alpha(t)}& \mbox{if $I=``\text{Level 2}"$.}
\end{array}
\right.
$$
%The dependence coefficients $\rho(t)$ and $\alpha(t)$ are defined in \eqref{rho_t_I} and \eqref{alpha_t_I}, respectively. 
The strength of the tail dependence varies from low (small values of $\rho(t)$, large values of $\alpha(t)$) to strong (large values of $\rho(t)$, small values of $\alpha(t)$) where $\eta(t,I) < 1$ unless $\rho(t) \equiv 1$ or $\alpha(t) \rightarrow 0$ (perfect dependence). 

For assessing the performance of our generalized additive modeling framework for the coefficient of tail dependence when different strengths of tail dependence are observed, we construct the following model for the angular dependence function: 
\begin{equation}
h\{\lambda_{\omegaV_0}(t,I)\} = \lambda_0 +s_1(t) \mathbbm{1}_{\lbrace I=``\text{Level 1}"\rbrace} + s_2(t) \mathbbm{1}_{\lbrace I=``\text{Level 2}"\rbrace}, \label{true_eta_simu_gauss_bevd_AI}
\end{equation}
with intercept $\lambda_0$ and smooth functions $s_1$ and $s_2$ of $t$. Our simulation study is based on an average of $100$ threshold exceedances of $M^{\min}_{\omegaV_0}=\min(2X^{E}_1,2X^{E}_2)$ for a fixed value of the covariate vector $(t,I)$; the sample size varies between $10^5$ for a threshold at the $90\%$ level and $10^6$ at the $99\%$ level. As before, $500$ repetitions are carried out. We consider different threshold levels to quantify the bias resulting from the estimation of $\eta(t,I)$ based on \eqref{M_tilde_AI} at a finite threshold $u$ or equivalently \eqref{AI_WT} at a finite level $x$. 

\begin{figure}[!h]
\begin{center}
\includegraphics[width=12cm,height=7cm]{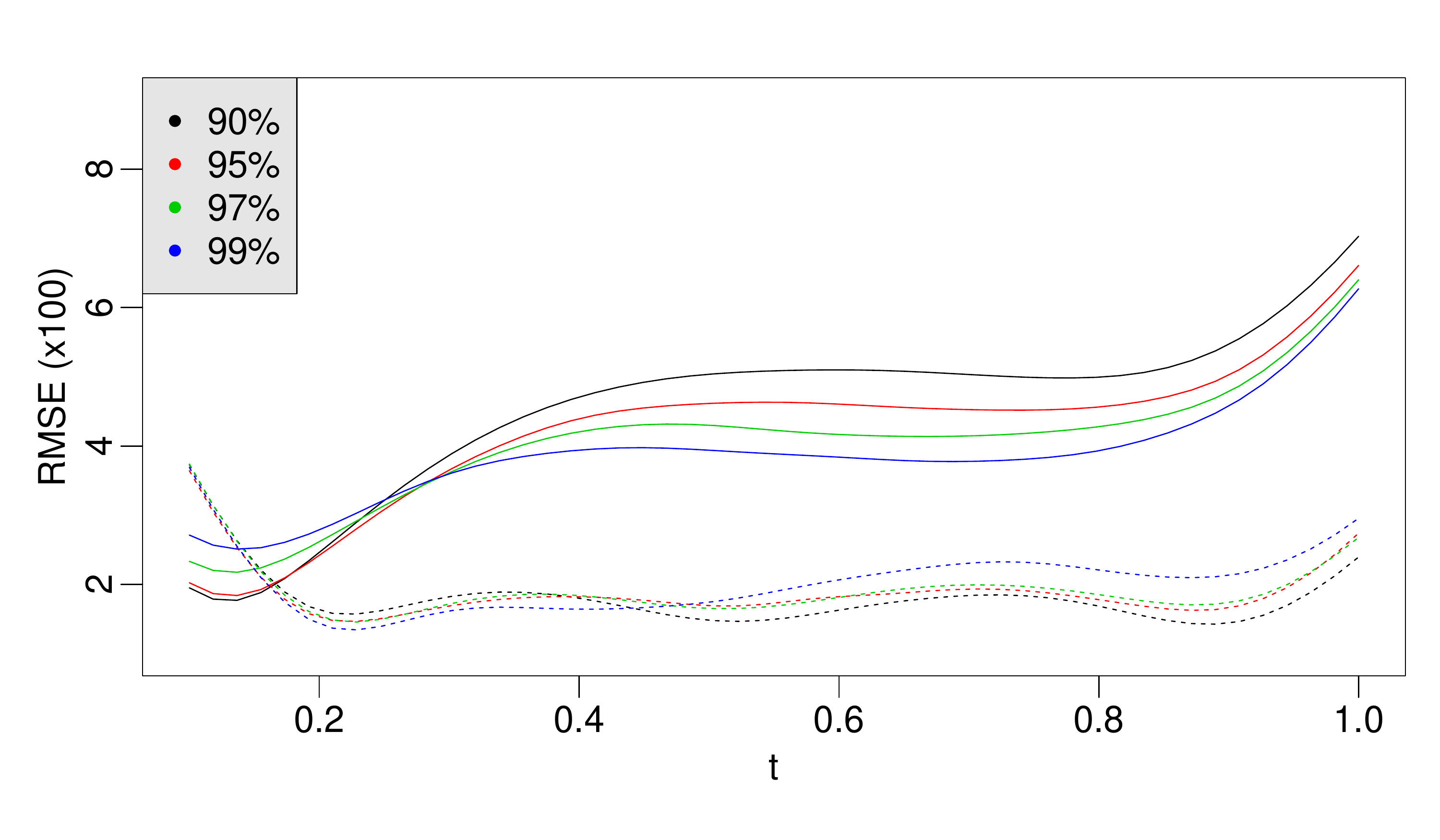}
  \caption{Root mean squared error of the bootstrap estimates of $\eta(t,I)$ for $I=1$, i.e., the Gaussian case (solid lines), and for $I=2$, i.e., the inverted logistic extreme value case (dashed lines). Different threshold levels are considered}
% Different threshold levels are considered: $90\%$ (black), $95\%$ (red), $97\%$ (blue), and $99\%$ (green).
\label{gauss_bevd_AI}
\end{center}
\end{figure}

%\subsubsection{Results}
The RMSE defined as 
\begin{equation}
\text{RMSE}(t,I)=\left[ \sum_{r=1}^{500} \{\hat{\eta}^{r}(t,I)-\eta(t,I)\}^2/500\right] ^{1/2}, \notag
\end{equation}
with $\hat{\eta}^{r}(t,I)$ the covariate-dependent tail dependence coefficient estimate obtained from the $r-$th bootstrap sample, is displayed in Figure \ref{gauss_bevd_AI}. In the Gaussian case, the RMSE decreases when the threshold level increases but increases when stronger dependence is considered, i.e., when $\rho(t)$ (or $t$, equivalently) increases. This is due to the slow convergence of tail measures at sub-asymptotic levels for the Gaussian dependence; see \cite{Coles1999}. In the inverted extreme value case, the estimator performance is largely unaffected by the threshold level as Equation \eqref{tail_IEVD} entails that the approximation \eqref{AI_WT} is exact at finite levels. 

%Overall, the bias seems to be unaffected by the choice of the threshold level. We observe however that by taking higher thresholds, the bias in the estimated coefficient of tail dependence in the Gaussian case ($I=\text{`Level 1'}$) with $\rho(t)>0.5$, gets slightly smaller. Since we managed to simulate exactly the same number of threshold excess at the different threshold levels and on average the same number of threshold excesses for the different values of $t$, the standard deviations are unsurprisingly unaffected by the threshold and dependence levels. Note that in the inverted extreme value case, the bias and the standard error are very similar at the different considered threshold levels which is explained by the fact that equation \eqref{tail_IEVD} holds, in contrast to \eqref{AI_WT}, for any $n\geq 1$.

\section{Application to nitrogen dioxide data}
\label{sec:appli}
\begin{figure}[!h]
	\centering
	\includegraphics[width=.75\textwidth]{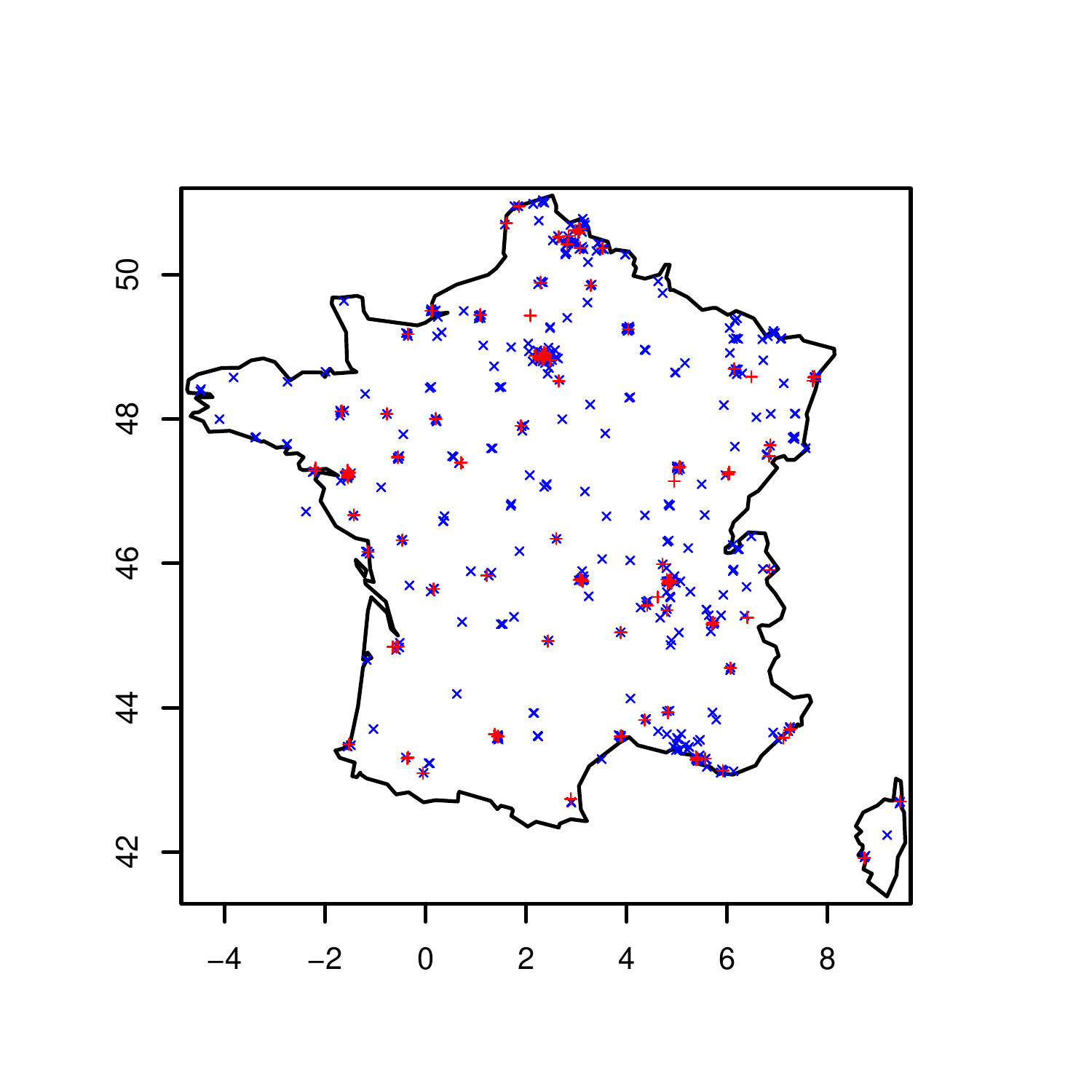}
	\caption{Map of nitrogen dioxide measurement stations in France that are ``background" (\textit{blue} $\times$ symbols) or ``traffic" (\textit{red} $+$ symbols) stations}
	\label{fig:map}
\end{figure}

\begin{figure}[!h]
 \centering 
 \textbf{Background stations} 
 \includegraphics[width=\textwidth]{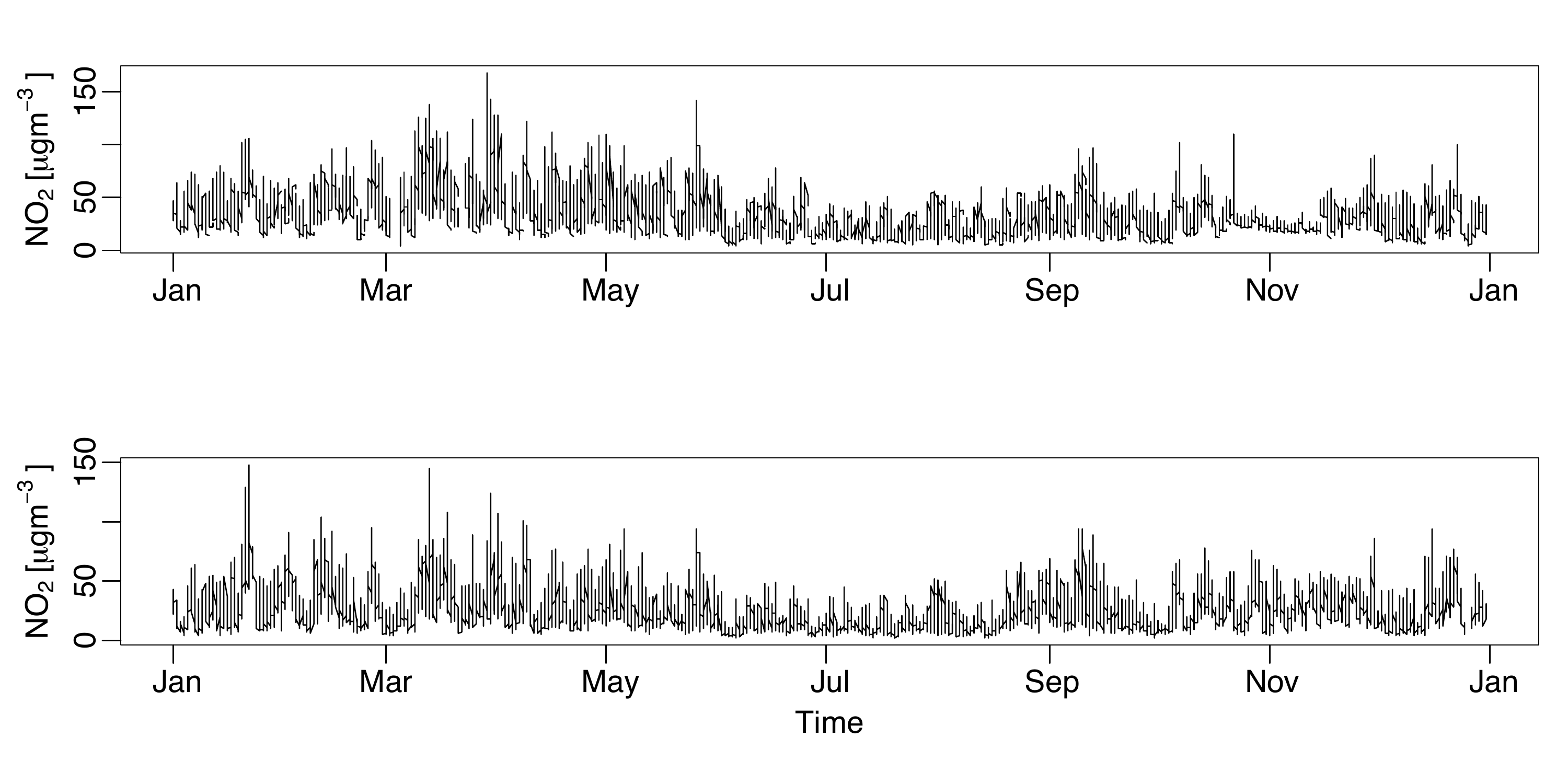}\\
 \textbf{Traffic stations} \\
 \includegraphics[width=\textwidth]{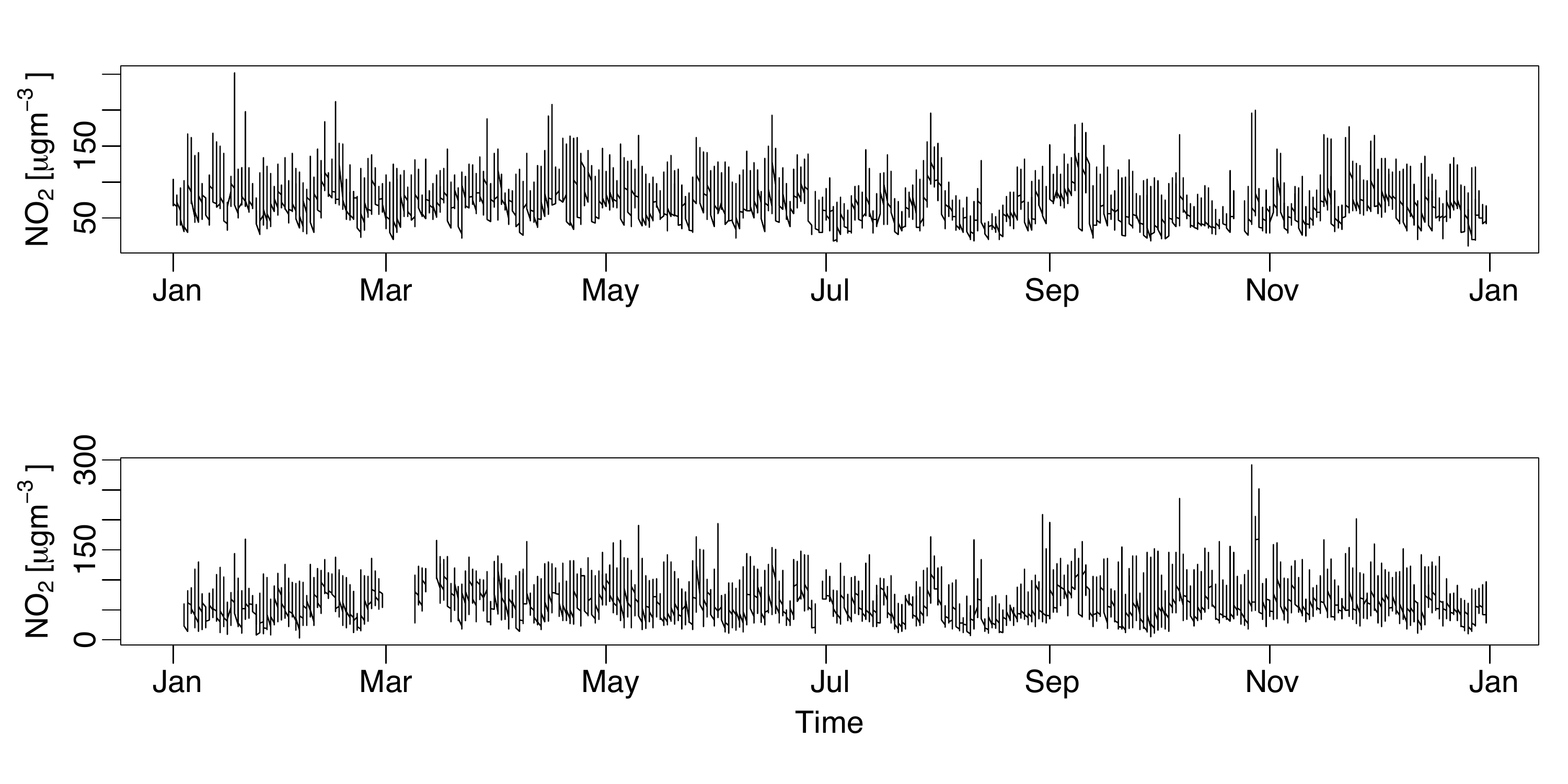}
 \caption{Nitrogen dioxide measurements recorded in $1999$ at two background stations in Metz (top) and two traffic stations in Paris (bottom)}
\label{data_background_traffic}
\end{figure}

We illustrate the modeling of covariate-dependent tail dependence on a nitrogen dioxide ($\text{NO}_2$) $[\mu\text{g/m}^{3}]$ data set, extracted from the European air quality database for pollutants AirBase\footnote{https://www.eea.europa.eu/data-and-maps/data/aqereporting-2}. It comprises $569$ measurement stations in France with hourly records of $\text{NO}_2$ observed over 14 years between 1999 and 2012; see the map of stations in Figure~\ref{fig:map}. With $\text{NO}_2$ produced mostly by the burning of fossil fuel and motor vehicle exhaust, we propose to distinguish traffic stations where the $\text{NO}_2$ level is predominantly determined by nearby traffic ($129$ stations) from background stations, often located in built-up areas, whose level of $\text{NO}_2$ is influenced by a combination of many sources ($440$ stations). Figure \ref{data_background_traffic} shows $\text{NO}_2$ measurements in $1999$ for two background stations located $3$ km apart (``Metz-Centre" and ``Metz-Borny") and for two traffic stations located $8$ km apart (``Auto A1-Saint-Denis" and ``Rue Bonaparte"). The measurements at the two background stations seem to follow a similar pattern, with large values recorded around the same time periods and low values observed during the summer season when most of the $\text{NO}_2$ is transformed into ozone through sunlight. The comparison of the $\text{NO}_2$ measurements for the two traffic stations is less straightforward. The magnitude of observations is much higher than for the background sites. Large observations seem to occur mostly locally although the stations being very close.

Such insights justify the distinction between the two station types when investigating the co-occurrence of large concentrations measured at pairs of stations. Moreover, it is natural to ask whether the frequency of co-occurrences of high pollution levels has changed over time, for instance, as a result of regulatory measures. For the present analysis, we reduce the dimension of data by considering monthly maxima at each of the $569$ stations, which avoids modeling of hourly patterns in $\text{NO}_2$ concentration levels and of intraday dependence between the measurements \citep{no2_daily_pattern}.

At each station, the marginal distribution of monthly maxima is modeled using a generalized extreme value (GEV) distribution. Its parameters for location $\mu$ and scale $\sigma$ are allowed to vary smoothly with the year and month of the observed maxima as follows: 
\begin{eqnarray}
\mu(t,m) &=& \mu_0 + f_1(t)+f_2(m), \notag \\
\sigma(t,m) &=& \sigma_0 + g_1(t) + g_2(m), \notag
\end{eqnarray}
where $t$ and $m$ denote the year and month of the observed maxima, respectively, $f_1$ and $g_1$ are smooth functions accounting for the trend, and $f_2$ and $g_2$ are cyclic smooth functions accounting for the seasonal component in the data \citep[Section 5.3]{wood_Book_2017}. Likelihood ratio tests are performed to assess whether we need smoothly varying terms or whether parametric and sinusoidal terms already provide a good fit. The final fitted model is then used to transform the data at each station to the unit Fr\'{e}chet distribution using the probability integral transform.

We focus on modeling the dependence between high $\text{NO}_2$ measurements recorded at pairs of stations. The dependence should naturally tend to decay with the distance between the stations in each pair. Therefore, we include distance (in kilometers) as a covariate in our model along with the type of area (traffic/background) and time (year), as discussed above. We consider the great circle distance between the stations, i.e., the shortest distance over Earth's surface, and fix the distance resolution at $10$ km, which represents the minimal distance by which stations must be separated to be distinguishable. This setting leads to $93$ distinct values for the distance covariate, $340,740$ pairs of observations for the background stations and $61,705$ pairs for the traffic stations.
Both tail measures, based on the assumption of either asymptotic dependence or asymptotic independence, are modeled. We construct the following ``full" models for the extremal coefficient and the tail dependence coefficient: 
\begin{eqnarray}
h\{\theta(t,d,\text{type})/2\} &=& \theta_0+\mathbbm{1}_{\lbrace \text{type}=``\text{Background}"\rbrace} \lbrace f_1(t)+f_2(d)+f(t,d)\rbrace + \notag \\
&&\mathbbm{1}_{\lbrace \text{type}=``\text{Traffic}"\rbrace} \lbrace \theta_1 + g_1(t)+g_2(d)+g(t,d)\rbrace, \label{theta_t_d} \\
h\left[ \{ 2\eta(t,d,\text{type})\}^{-1}\right] &=& \eta_0+\mathbbm{1}_{\lbrace \text{type}=``\text{Background}"\rbrace} \lbrace \tilde{f}_1(t)+\tilde{f}_2(d)+\tilde{f}(t,d)\rbrace + \notag \\
&&\mathbbm{1}_{\lbrace \text{type}=``\text{Traffic}"\rbrace} \lbrace \eta_1 + \tilde{g}_1(t)+\tilde{g}_2(d)+\tilde{g}(t,d)\rbrace, \label{eta_t_d}
\end{eqnarray}
where $t$ represents time (in years), $d$ the distance between the stations in each pair (in km), and $h(x)=\log(x-1/2)-\log(1-x)$. The interaction between time and distance is represented using a tensor product basis \citep[Section 5.6]{wood_Book_2017}. We fit models \eqref{theta_t_d} and \eqref{eta_t_d} based on the deficits of the $5\%$ quantile of $M^{\min\downarrow}_{\omegaV_0}$ and the exceedances of the $95\%$ quantile of $M^{\min}_{\omegaV_0}$ with $\omegaV_0=(0.5,0.5)$, respectively. 

\begin{figure}[!h]
	\centering 
 \subfloat{\includegraphics[width=0.5\textwidth]{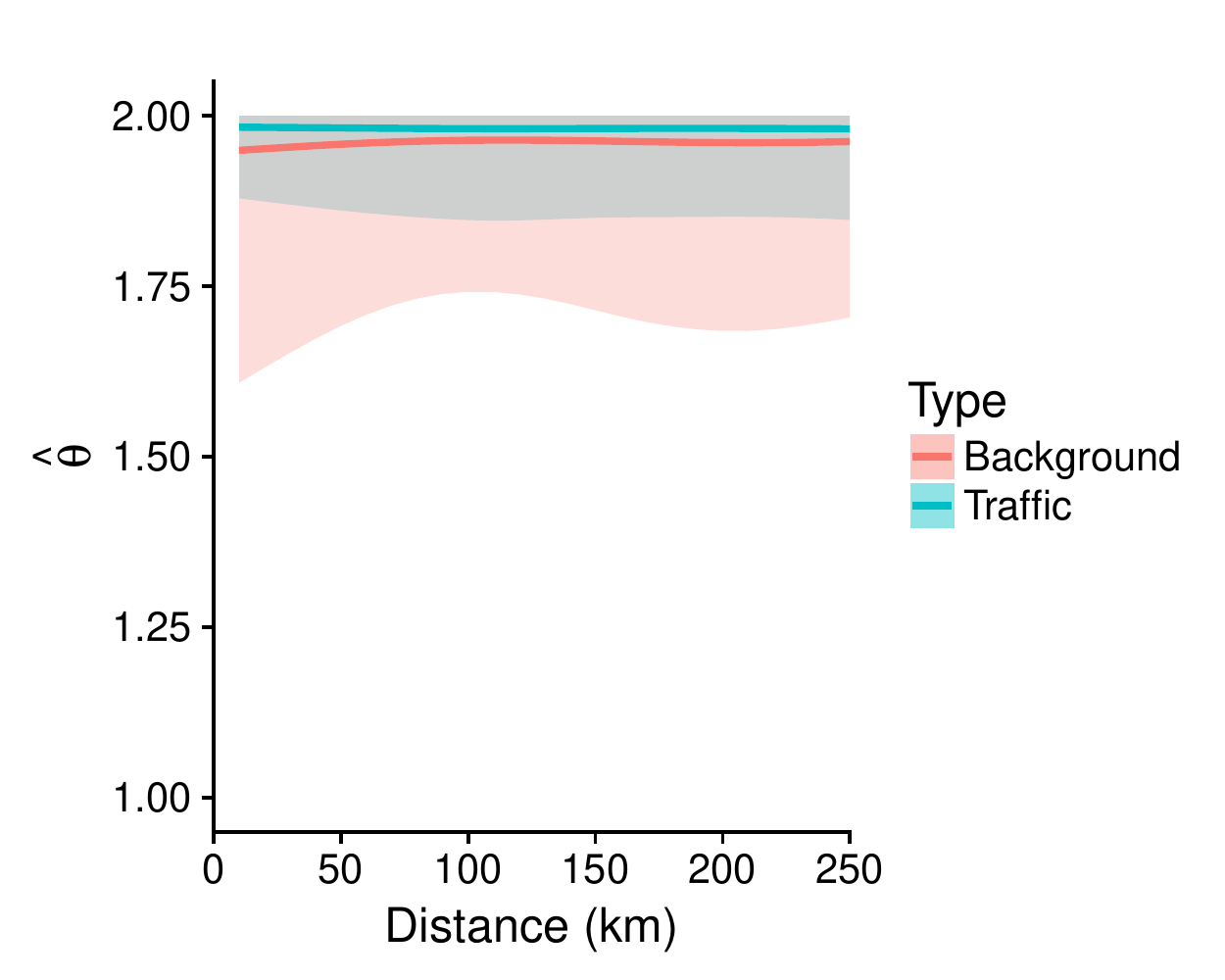}} \hfill
 \subfloat{\includegraphics[width=0.5\textwidth]{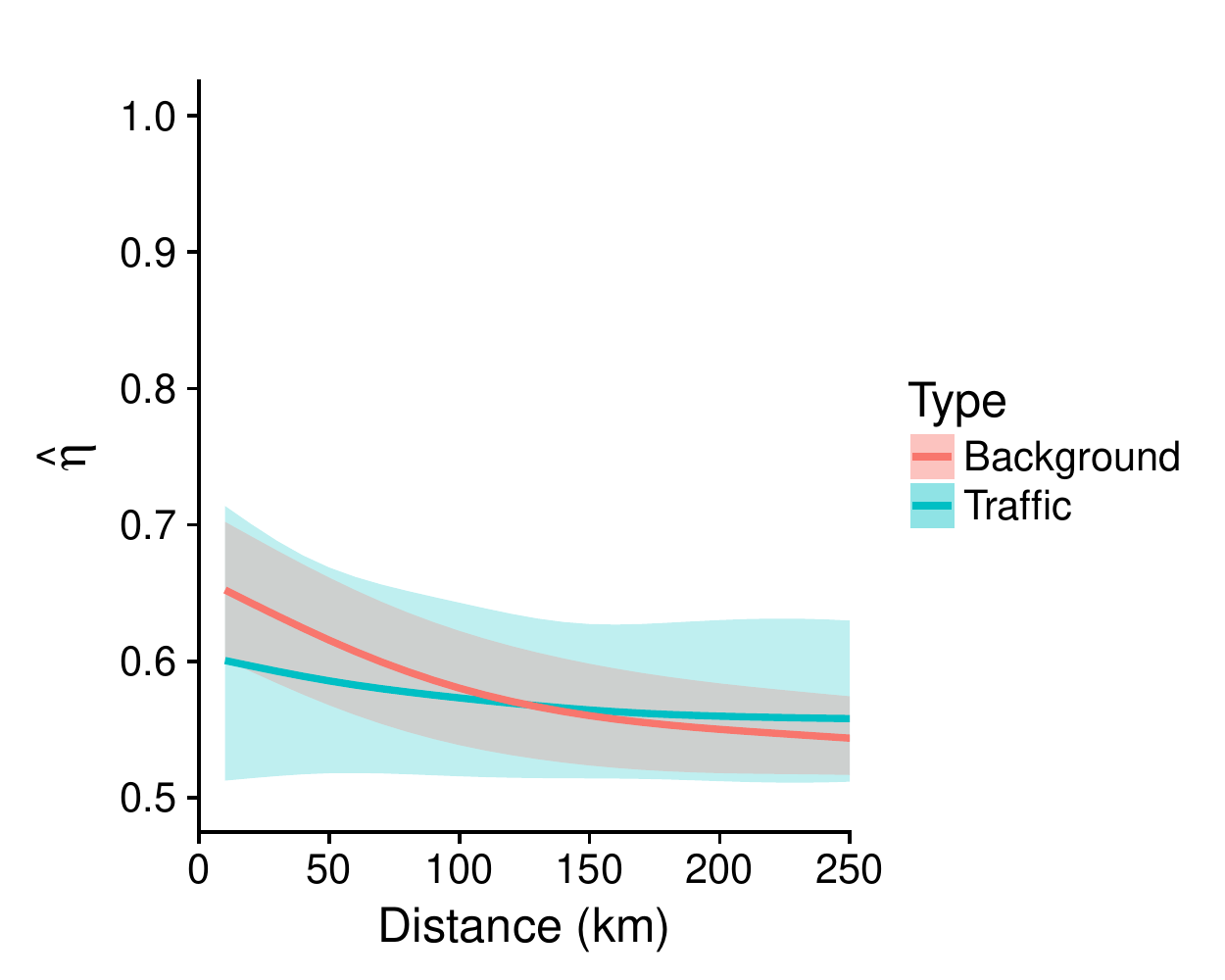}}
	\caption{Estimation of distance-dependent dependence summaries for the nitrogen dioxide data. Left: $\hat{\theta}(d,\text{type})$, assuming asymptotic dependence. Right: $\hat{\eta}(d,\text{type})$, assuming asymptotic independence}
	\label{fig:distonly}
\end{figure}

To start, we conduct a simpler, purely spatial analysis and consider the models without time effects and such that data are pooled together over the whole period for each pair of stations during the estimation. Estimated summaries $\hat{\theta}(d,\text{type})$ and $\hat{\eta}(d,\text{type})$ are shown in Figure~\ref{fig:distonly}; they clearly hint at asymptotic independence with estimates and bootstrap-based pointwise confidence intervals of $\theta$ very close to $2$, while those of $\eta$ are clearly bounded away from $1$. Overall, the joint tail decay rates in asymptotic independence appear to be quite fast with pointwise confidence envelopes of $\eta$ contained between $0.5$ and $0.7$ approximately. A partial explanation for this relatively weak dependence is that our univariate models have already appropriately removed seasonal trends in the data, such that dependence in the resulting residuals cannot arise from intermediate-range clustering in space and time. As a matter of fact, as seasonal patterns are typically spatial, filtering out these patterns would result in a spatial de-clustering in the region where the seasonal features are observed. We detect a stronger weakening of dependence with increasing distance in the background stations, which makes sense because the peaks in the $\text{NO}_2$ concentrations are strongly influenced by nearby traffic. 
%\CV{On peu ajouter quelque chose comme: Indeed seasonal trend is typically spatial so that removing it prevents to get similar seasonal pattern for stations in the same region}. 

Next, we consider regression fits~\eqref{theta_t_d} and \eqref{theta_t_d} with spatial and temporal components; i.e., we have a more complex, higher-dimensional model for the predictors, and we expect the estimation uncertainty to be higher. Table~\ref{summary_gam} summarizes both fitted models. All the considered covariate effects, except for the distance between traffic stations when the tail dependence coefficient is modeled, are statistically significant. 

\begin{table}[!h]
\caption{Estimates (se) of the intercepts and the degrees of freedom (edf) of the smooth functions in models~\eqref{theta_t_d} and \eqref{eta_t_d}}
\label{summary_gam}
\centering
\begin{scriptsize}
\begin{tabular}{cclc|c|c|c|c|c|c}
\cline{3-10}
& \multicolumn{1}{c|}{}  & \multicolumn{1}{c|}{$\hat{\theta}_0$} & $\hat{\theta}_1$ & $\hat{f}_1$   & \multicolumn{1}{c|}{$\hat{g}_1$} & \multicolumn{1}{c|}{$\hat{f}_2$}  & \multicolumn{1}{c|}{$\hat{g}_2$}  & \multicolumn{1}{c|}{$\hat{f}$}              & \multicolumn{1}{c|}{$\hat{g}$}   \\[5pt] \hline
\multicolumn{1}{|c|}{\multirow{3}{*}{$\hat{\theta}(t,d,\text{type})$}} & \multicolumn{1}{|c|}{estimate/edf} & \multicolumn{1}{|c|}{$12.76 (0.39)$}   & $-2.41 (0.69)$ & $3.98$  & \multicolumn{1}{c|}{$3.55$} & \multicolumn{1}{c|}{$3.82$}  & \multicolumn{1}{c|}{$0.84$} & \multicolumn{1}{c|}{$15.55$} & \multicolumn{1}{c|}{$2.92$} \\[5pt] \cline{2-10}
\multicolumn{1}{|c|}{} & \multicolumn{1}{|c|}{$p$-value}   & \multicolumn{1}{|c|}{$< 10^{-16}$} & $4.74 \times 10^{-4}$ & $< 10^{-16}$ & $< 10^{-16}$ & $< 10^{-16}$  & $1.21 \times 10^{-2}$ & $< 10^{-16}$ & \multicolumn{1}{c|}{$< 10^{-16}$}  \\[5pt] \hline
%\end{tabular}
%\end{scriptsize}
%\bigskip
%\centering
%\begin{scriptsize}
%\begin{tabular}{cclc|c|c|c|c|c|c}
\cline{3-10}
& \multicolumn{1}{c|}{}  & \multicolumn{1}{c|}{$\hat{\eta}_0$} & $\hat{\eta}_1$ & $\hat{\tilde{f}}_1$   & \multicolumn{1}{c|}{$\hat{\tilde{g}}_1$} & \multicolumn{1}{c|}{$\hat{\tilde{f}}_2$}  & \multicolumn{1}{c|}{$\hat{\tilde{g}}_2$}  & \multicolumn{1}{c|}{$\hat{\tilde{f}}$}   & \multicolumn{1}{c|}{$\hat{\tilde{g}}$}   \\[5pt] \hline
\multicolumn{1}{|c|}{\multirow{3}{*}{$\hat{\eta}(t,d,\text{type})$}} & \multicolumn{1}{|c|}{estimate/edf} & \multicolumn{1}{|c|}{$1.86 (0.03)$}   & $-0.46 (0.07)$ & \multicolumn{1}{c|}{$3.88$} & \multicolumn{1}{c|}{$3.11$}  & \multicolumn{1}{c|}{$3.82$} & $0$ & \multicolumn{1}{c|}{$9.62$} & \multicolumn{1}{c|}{$1.66$} \\[5pt] \cline{2-10}
\multicolumn{1}{|c|}{} & \multicolumn{1}{|c|}{$p$-value}   & \multicolumn{1}{|c|}{$< 10^{-16}$} & $4.49 \times 10^{-10}$ & $< 10^{-16}$ & $< 10^{-16}$ & $< 10^{-16}$  & $0.61$ & $< 10^{-16}$ & \multicolumn{1}{c|}{$2.98 \times 10^{-4}$}  \\[5pt] \hline
\end{tabular}
\end{scriptsize}
\end{table}

\begin{figure}[!h]
 \centering 
 \subfloat{\includegraphics[width=0.5\textwidth]{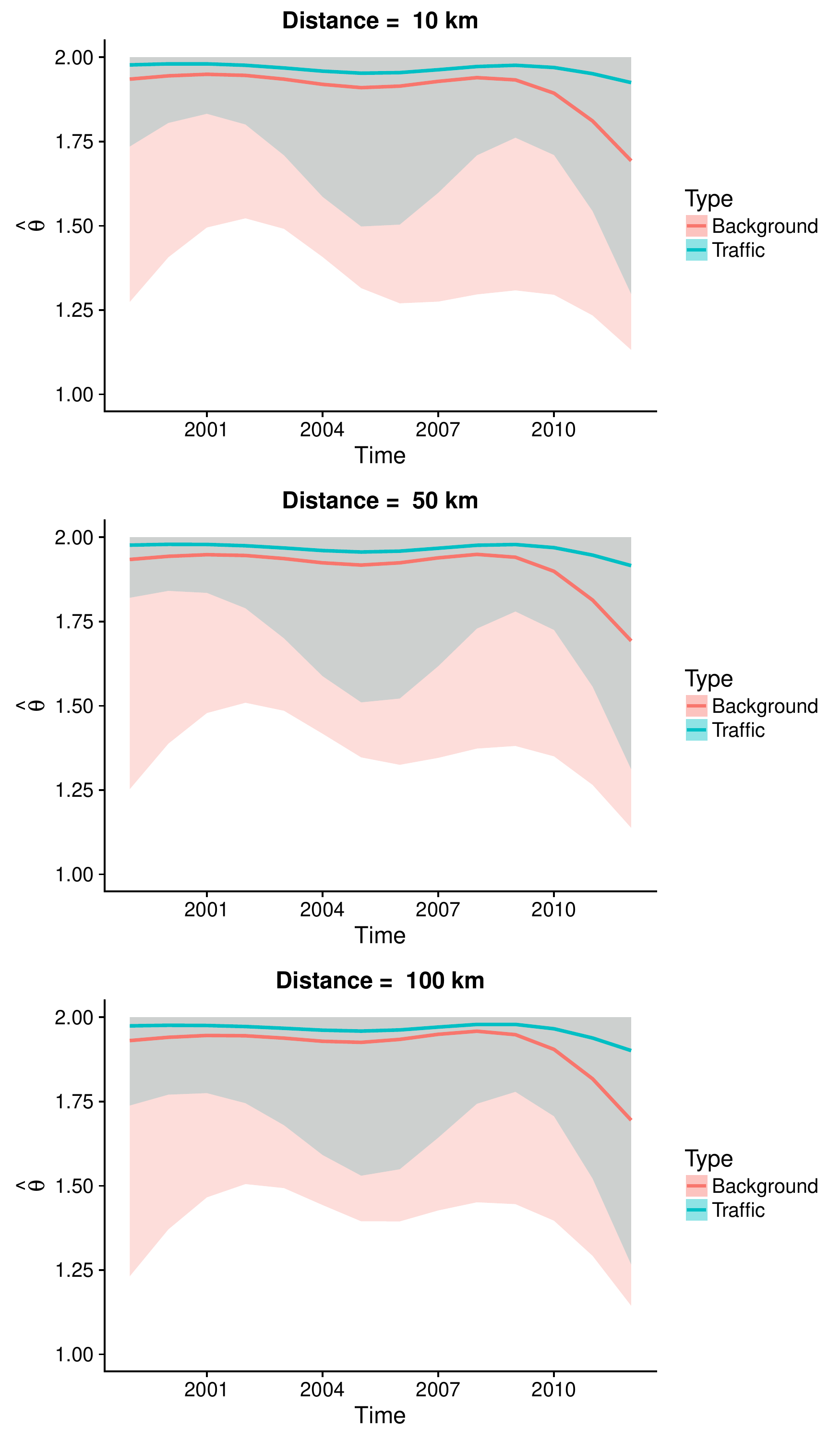}} \hfill
 \subfloat{\includegraphics[width=0.5\textwidth]{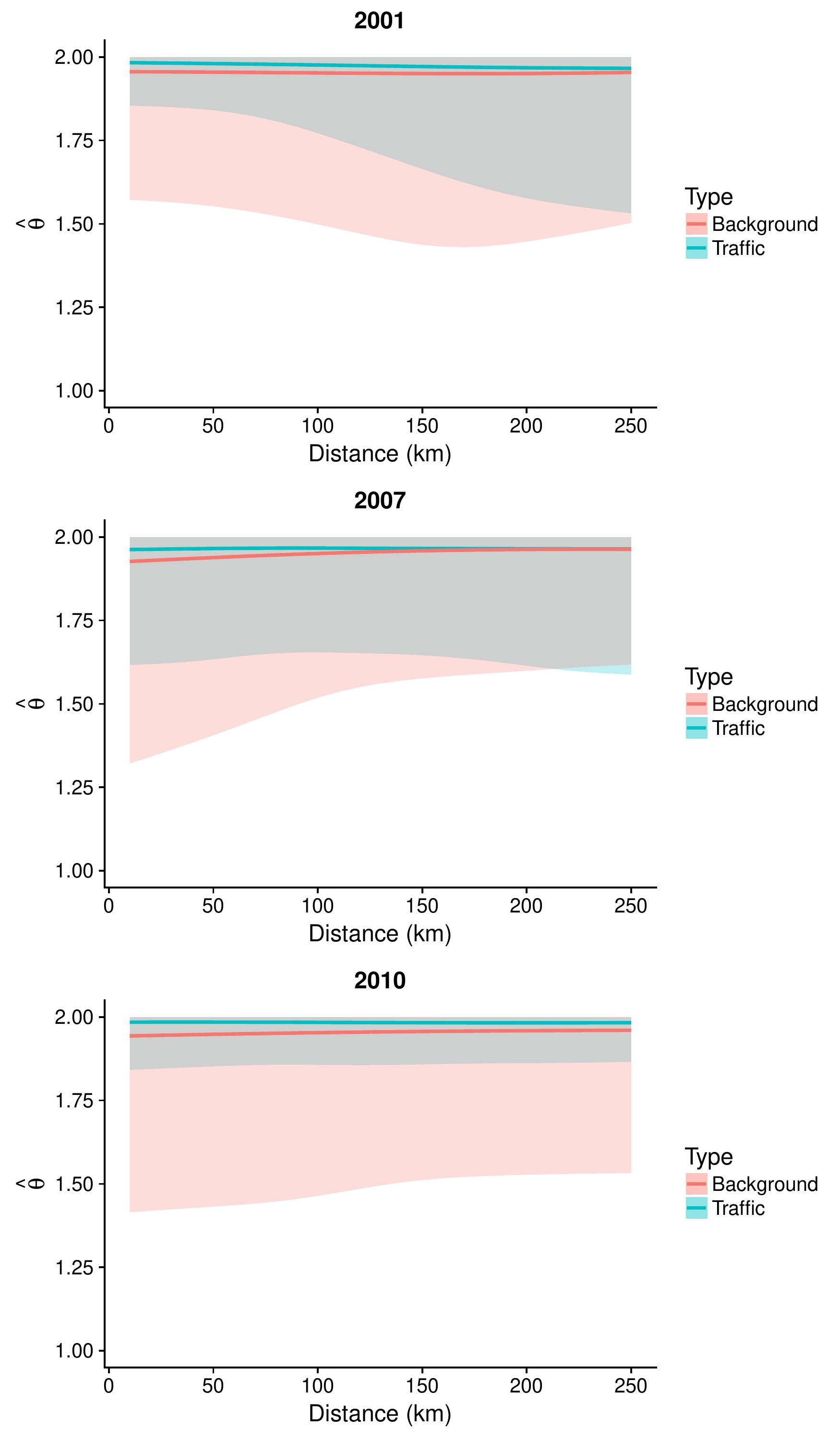}}
 \caption{Cross sections of the covariate-dependent $\hat{\theta}(t,d)$ with the corresponding $95\%$ bootstrap uncertainty intervals. The bootstrap is based on $300$ bootstrap samples}
\label{theta_full_model}
\end{figure}

\begin{figure}[!h]
 \centering 
 \includegraphics[width=0.47\textwidth]{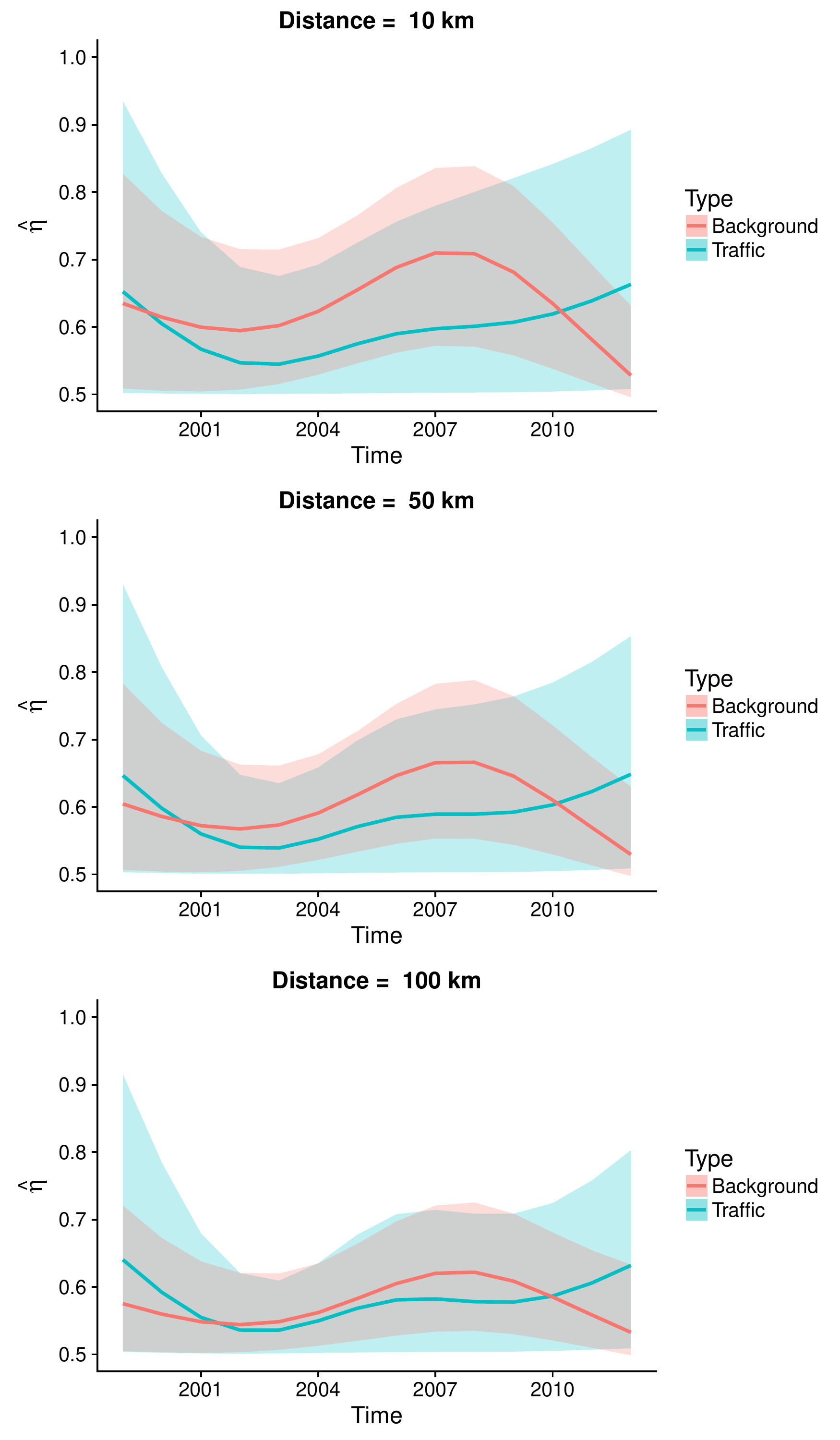}
 \includegraphics[width=0.47\textwidth]{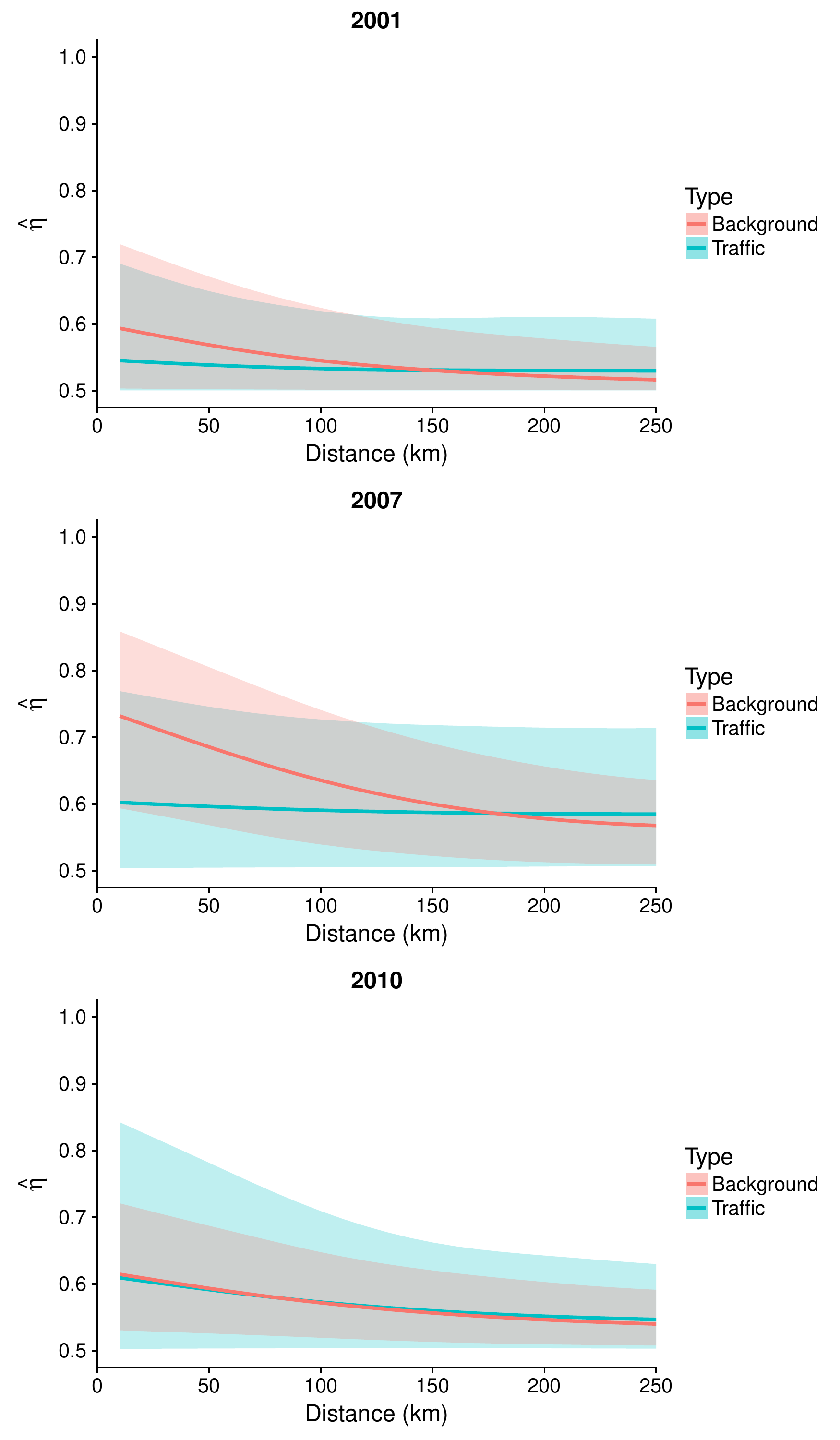}
 \caption{Cross sections of the mean bootstrap estimate of $\hat{\eta}(t,d)$ along with the corresponding $95\%$ bootstrap uncertainty intervals. The bootstrap procedure is based on $300$ bootstrap samples}
\label{eta_full_model}
\end{figure}

\noindent This confirms our intuition that high $\text{NO}_2$ concentrations are relatively localized for the traffic stations. A block bootstrap procedure treating $\text{NO}_2$ measurements from each month and each year as independent for each type of area is used to assess the uncertainty. Figure~\ref{theta_full_model} shows cross sections of the estimates of the extremal coefficient for the traffic and background stations; we fix one of the two continuous covariates to show the smooth effect of the other. 
For both area types, the estimates are very close to the upper bound of the extremal coefficient for almost all values of $t$ and $d$. This result implies weak extremal dependence between the $\text{NO}_2$ measurements at the pairs of stations, and we can suppose asymptotic independence in the data. We now focus on the estimates of the tail dependence coefficient; the sub-asymptotic modeling of the tails should be more informative in this case. Figure~\ref{eta_full_model} displays cross sections of the estimate of the tail dependence coefficient $\hat{\eta}(t,d)$ for the different types of area.
For the considered years, the residual tail dependence is relatively weak and largely unaffected by the distance between traffic stations, owing to the very localized features of traffic pollution inducing high concentrations of $\text{NO}_2$. Unreported results with a distance resolution of $5$ km have shown globally higher tail dependence estimates (over time) for close traffic stations that are at most $5$ km apart. The uncertainty in the estimators $\hat{\eta}(t,d,\text{``Traffic"})$ is relatively high due to the small amount of information available from the traffic stations. For background stations, the smooth effect of distance on the tail dependence is more pronounced with a decrease in the dependence at larger distances. The sharpness of the decrease shows some variation over time, with higher tail dependence for close background stations observed in $2007$. The smooth effect of time on the tail dependence is important mainly for small distances with an overall slight increase over time for the traffic stations and a bump (increase) in the dependence around $2005$ and $2009$ for the background stations. These effects are observable for stations up to $100$ km apart although with a lower magnitude of the dependence measure.

\section{Conclusion}
\label{sec:conclusion}
Starting from a $d$-dimensional random vector $\mathbf{X}$ with either asymptotic dependence or independence, we developed min- and max-projection techniques allowing us to simplify the joint tail characterization problem to univariate modeling with well-understood exponential distributions for which generalized additive modeling under censoring is feasible and well-known from survival modeling. The exponential rate carries crucial information about the form and the rate of the joint tail decay in different directions. This setup facilitates flexible inference for the tail dependence as it is based on the excesses or deficits of a univariate exponential random variable while censoring observations that do not contribute to the joint tail, and it allows us to include multiple covariates of different types through the GAM framework. Although we focused on estimating the covariate influence for a fixed direction, we can apply the projection technique of \cite{Mhalla2017} in different directions to obtain smooth and valid estimates of the Pickands dependence function or the angular dependence function under shape constraints and for a fixed set of covariates. 

Our application demonstrates that it is useful to apply both projection techniques in practice to compare covariate-driven estimates for tail dependence summaries in each of the two asymptotic regimes. Our pairwise modeling of $\text{NO}_2$ measurements in France to investigate the effect of time and spatial distance on the joint tail behavior showed strong evidence against asymptotic dependence. The results of our application gave strong support for asymptotic independence with estimated extremal coefficients close to $2$ and the confidence intervals of tail dependence coefficients bounded away from $1$. Our methodology constitutes an important step toward the distinction between asymptotic dependence and independence, when there is no clear evidence for one of the two models or when different model classes arise for different covariate configurations. Formal hypothesis testing of asymptotic dependence against asymptotic independence for a fixed set of covariates would be an important extension. As our censoring mechanisms select different observations according to the two models, likelihood-based tests are not directly applicable, but non-parametric tests of extremal dependence could provide guidance \citep[e.g.,][Chapters 17 and 18]{dey2015extreme}. Finally, depending on the application context, a threshold that varies with covariates and/or directions could be used to determine excesses and deficits, which could reduce or homogenize estimation bias and uncertainty. 
\newpage
\bibliographystyle{agsm}
\bibliography{mybib}
\end{document}